\documentclass[manuscript]{aastex}
\makeindex
\makeatletter

\usepackage{natbib}
\usepackage{multirow}

\def \bvec{\bf{B}}

\def \vlos{\mathbf {V_{LOS}}}
\def \evec{\bf{E}}

\def \spot{{S$_{z,pot}$}}

\def \sf{{S$_{z,free}$}}

\def\gsim{\,\lower3pt\hbox{$\sim$}\llap{\raise2pt\hbox{$>$}}\,}
\def\lsim{\,\lower3pt\hbox{$\sim$}\llap{\raise2pt\hbox{$<$}}\,}

\newcommand{\Rsun}{\ensuremath{R_{\odot}}}
\newcommand{\vecB}{{\mathbf B}}
\newcommand{\vecE}{{\mathbf E}}

\newcommand{\vecv}{{\mathbf V}}

\newcommand{\vecbhat}{{\mathbf {\hat b}}}

\newcommand{\veclhat}{{\mathbf {\hat l}}}
\newcommand{\veczhat}{{\mathbf {\hat z}}}
\newcommand{\vecqhat}{{\mathbf {\hat q}}}
\newcommand{\scrJ}{{\mathcal{J}}}
\newcommand{\scrB}{{\mathcal{B}}}

\newcommand{\be}{\begin{equation}}
\newcommand{\ee}{\end{equation}}

\newcommand{\bex}{\begin{equation}\notag}
\newcommand{\eex}{\end{equation}\notag}
\newcommand{\bea}{\begin{eqnarray}}
\newcommand{\eea}{\end{eqnarray}}

\shorttitle{} \shortauthors{Kazachenko et. al}
\def\Rsun{\ifmmode{R_\odot}\else{R$_\odot$}\fi}

\begin{document}

\title{A Comprehensive Method of Estimating Electric Fields from Vector Magnetic Field and Doppler Measurements}

\author{Maria D. Kazachenko\altaffilmark{1}, George H. Fisher\altaffilmark{1}, Brian T. Welsch\altaffilmark{1}}

\altaffiltext{1}{Space Sciences Laboratory, UC Berkeley, CA 94720}

\email{kazachenko@ssl.berkeley.edu}

\begin{abstract}
Photospheric electric fields, estimated from sequences of vector
magnetic field and Doppler measurements, can be used to estimate the
flux of magnetic energy (the Poynting flux) into the corona and as
time-dependent boundary conditions for dynamic models of the coronal
magnetic field.  We have modified and extended an existing method to
estimate photospheric electric fields that combines a
poloidal-toroidal (PTD) decomposition of the evolving magnetic field
vector with Doppler and horizontal plasma velocities.  Our current,
more comprehensive method, which we dub the ``{\bf P}TD-{\bf
 D}oppler-{\bf F}LCT {\bf I}deal'' (PDFI) technique, can now
incorporate Doppler velocities from non-normal viewing angles. It uses
the \texttt{FISHPACK} software package to solve several
two-dimensional Poisson equations, a faster and more robust approach
than our previous implementations.  Here, we describe systematic,
quantitative tests of the accuracy and robustness of the PDFI
technique using synthetic data from anelastic MHD (\texttt{ANMHD})
simulations, which have been used in similar tests in the past. We
find that the PDFI method has less than $1\%$ error in the total
Poynting flux and a $10\%$ error in the helicity flux rate at a normal
viewing angle $(\theta=0$) and less than $25\%$ and $10\%$ errors
respectively at large viewing angles ($\theta<60^\circ$). We compare
our results with other inversion methods at zero viewing angle, and
find that our method's estimates of the fluxes of magnetic energy and
helicity are comparable to or more accurate than other methods.  We
also discuss the limitations of the PDFI method and its uncertainties.
\end{abstract}

\onecolumn
\thispagestyle{empty}

\tableofcontents

\newpage
\pagenumbering{arabic}

\section{INTRODUCTION}
\label{intro}

Energy from magnetic fields on the Sun powers nearly all
manifestations of solar activity, from heating in active regions and
the solar wind, to dramatic events like flares and coronal mass
ejections (CMEs).  Quantitative studies of the flow of magnetic energy
through the solar atmosphere require a knowledge of both the magnetic
and electric field vectors.  The goal of this paper is to show how
time sequences of vector magnetic field maps (vector magnetograms),
along with other observational constraints such as Doppler flow
measurements, can be used in a practical way to construct maps of the
electric field vector.  Knowing both the electric and magnetic field,
we can find the Poynting flux of electromagnetic energy and the flux
of relative magnetic helicity, important quantities that describe how
magnetic energy is transported, stored, and released in the solar
atmosphere.

Past attempts to determine the electric field distribution on the Sun
have followed two approaches: (1) direct spectroscopic measurements of
the electric field using the linear Stark effect, (2) indirect
determinations of the electric field by using Faraday's law, relating
the temporal derivative of the magnetic field to the curl of the
electric field.  \cite{Wien1916} was the first to suggest measuring
the electric field of solar plasma directly using the Stark effect. In
the 1980s, some attempts were made to measure the electric field at
the Sun using helium and silicon spectra, finding an electric field of
$700$ V/cm \citep{Davis1977} and $300$ V/cm \citep{Jordan1980}.
Later, examining Stark-broadened Paschen emission lines in hydrogen,
\cite{Moran1991} found an upper limit of $5-10$ V/cm.  In the same
work they pointed out that the direct measurement of the electric
field using the Stark effect is difficult because of the low sensitivity
of the measurements.

Indirect determinations of the electric field have been considerably
more successful.  The implementation of local-correlation tracking
techniques ({\texttt LCT}, e.g. \citealt{November1988, Title1995,
  Hurlburt1995, Fisher2008}), applied to time-sequences of
magnetograms allows one to determine a ``pattern motion'' velocity of
the line-of-sight component of the magnetic field in the plane of the
magnetogram \citep{Demoulin2003}.  Interpreting the pattern motion as a plasma velocity,
and assuming an ideal MHD Ohm's law for the electric field ($c {\bf E}
= - {\bf V} \times {\bf B} $), allows one to determine some of the
information about the velocity field, and hence by assumption, the
electric field.  These techniques can be improved by explicitly
incorporating the normal component of the magnetic induction equation
into the solution for the components of the velocity ${\bf V}$
\citep{Kusano2002,Welsch2004,Longcope2004c,Schuck2006,Schuck2008,Chae2008},
resulting in improvements in the velocity inversion
\citep{Welsch2007}.

More recently, we have developed inversion techniques for the electric field
itself rather than the velocity field, based on a ``Poloidal-Toroidal Decomposition'' (PTD) of the magnetic field
and its time derivative.  These techniques use Faraday's law and 
other theoretical and observational constraints to determine the electric
field 
\citep{Fisher2010,Fisher2012}.  Inversions for the electric field
instead of the velocity field offer a few distinct advantages.  
First, the PTD method incorporates additional information from
evolution of the horizontal magnetic field (the field parallel to the
photosphere).  This is used to make an estimate of the normal electric
field, independent of the horizontal electric field inferred by
methods that derive ${\bf V}$ from the normal component of the
induction equation.
Second, in regions where the magnetic field is relatively weak or
uncertain, the determination of the velocity field is especially
ill-posed:
for instance, outside of active regions, fluctuations in the normal
magnetic field can be dominated by noise, resulting in a wildly
varying and unphysical behavior in the inferred ${\bf V}$.  Complex
masking and filtering techniques must be applied \textit{post facto}
to suppress this behavior.
The solutions for the electric field, on the other hand, may 
contain small errors in regions with small magnetic field strength,
but the solutions vary smoothly within regions containing both strong and weak 
magnetic fields.
Third, if a model for non-ideal effects is included in the electric
field ${\bf E}$ assumed to drive the change in ${\bf B}$, they can be
captured by the PTD solutions, while the velocity formalism may give
spurious results.

\cite{Fisher2010} described in detail the PTD techniques necessary to
find the ``inductive'' electric field that satisfies Faraday's law,
given the spatial distribution of the temporal derivative of the
magnetic field measured across a region of the solar photosphere.
They also noted that the resulting solutions are not unique, in that
gradients of arbitrary scalar functions could be added to the PTD
solutions without affecting Faraday's law.  They presented several
approaches to specify the electric field uniquely, including requiring
the electric field to be perpendicular to the magnetic field,
consistent with the ideal Ohm's law. \cite{Fisher2012} demonstrated
that by adding to the PTD solutions the gradients of scalar functions
that are constructed to match the electric field near polarity
inversion lines (PILs), the accuracy of electric field reconstructions
could be substantially improved, beyond all of the velocity inversion
techniques considered in the comparative study of \cite{Welsch2007}.
To match the electric field near PILs, one can use measurements of the
Doppler velocity and the transverse magnetic field distribution in
those regions.

In this paper, we build on \cite{Fisher2012} results in a number of
ways, with the goal of carefully describing and validating a practical
implementation of the introduced methods that could be used to
routinely find the electric field from sequences of vector magnetogram
observations, such as those available from the Helioseismic and
Magnetic Imager (HMI) \citep{Schou2012,Scherrer2012} on NASA's SDO
mission.  First, we adopt the use of the \texttt{FISHPACK} software
library for Helmholtz equations \citep{Schwarztrauber1975}, developed
at NCAR in the 1970s, for solving the two-dimensional Poisson
equations that are at the heart of the PTD inversion technique.  This
software, based on the cyclic reduction technique, is very efficient,
and explicitly allows for the specification of normal-derivative
(Neumann) boundary conditions that are consistent with our desired
boundary conditions on the electric field at the edges of vector
magnetogram images.  Second, we develop the ability to incorporate
line-of-sight Doppler measurements that are taken at non-normal
viewing angles, appropriate for vector magnetograms that are not at
disk center, to determine electric fields near PILs.  Finally, to
validate our inversion methods, we analyze the performance of our
techniques, comparing electric field inversions with a test case from
an MHD simulation where the true electric field is known
\citep{Welsch2007}.  Besides the electric field, we compare
derived values of the Poynting flux and the relative helicity fluxes
with the known values from the test case.  Finally, in
Appendix~\ref{sphere} we describe how we adapt our electric field
solution techniques to spherical polar coordinates, appropriate for
larger fields of view on the Sun.

The remainder of the paper is structured as follows.  In
\S~\ref{ptdtheory} we describe the electric field inversion technique,
starting from the basic PTD formalism, and also improvements that
account for a non-zero viewing angle. In \S~\ref{poynting-hel} we
describe how to calculate Poynting and helicity fluxes, using our
derived electric field. This includes a decomposition of the Poynting
flux into fluxes of potential and free magnetic energies. In
\S~\ref{anmhd} we perform tests of the electric field inversion method
using the \texttt{ANMHD} simulation test case, and provide validation
metrics that quantify how well the inversion recovers the true
solution.  Finally, in \S~\ref{disc}, we summarize the strengths and
weaknesses of our technique and describe how it will be used to
analyze HMI vector magnetograms.

\section{FINDING ELECTRIC FIELDS}\label{ptdtheory}
\subsection{The Inductive PTD Electric Field: $\bf{E^P}$}\label{PTD}

Poloidal-toroidal decomposition (PTD) allows one to estimate the photospheric electric field vector from the evolution of the photospheric magnetic field vector. 
Here we present a brief synopsis of the PTD 
method \citep{Fisher2010}, plus improvements we have made since that article
was published.

The fundamental idea of PTD is that the magnetic field vector 
${\bvec}=(B_x,B_y,B_z)$ has a solenoidal nature and hence can be specified
by two scalar functions,  the ``poloidal'' and  ``toroidal''  potentials  
\citep{Chandrasekhar1961,Moffatt1978},
which we denote
$\mathcal{B}$ and $\mathcal{J}$ following the notation used in descriptions
of anelastic MHD which employ these potentials \citep{Lantz1999}:
 
\begin{equation}
 {\bvec}=\nabla \times \nabla \times \mathcal{B}  
{\bf \hat{z}} + \nabla \times \mathcal{J}  {\bf \hat{z}}.
\label{eq1}
\end{equation}
Here we consider a locally Cartesian coordinate system $(x,y,z)$, that has 
its $z$ axis oriented perpendicular to the photosphere, with the positive direction away from the Sun's
center (see  Figure~\ref{eyeball}; the spherical case is described in Appendix~\ref{sphere}). We 
use subscript $z$ to denote vector components or derivatives in the vertical
direction, and
subscript $h$ to denote vector components or derivatives in the 
horizontal directions, i.e. parallel to the photosphere. 

Taking the partial time derivative of Eq.~(\ref{eq1}), one finds 
\begin{equation}
\dot{ {\bvec}}=\nabla \times \nabla \times \dot{\mathcal{B}} 
{\bf \hat{z}} + \nabla \times \dot{\mathcal{J}} {\bf \hat{z}}.
\label{eq2}
\end{equation}
By
examining the $z$-component of Eq.~(\ref{eq2}), its horizontal divergence,
and the z-component of its curl,
we find three two-dimensional Poisson 
equations for the unknown functions $\dot{\mathcal{B}}$,  
$\frac{\partial \dot \mathcal{B}}{\partial z}$ and  $\dot{\mathcal{J}}$
in terms of known physical quantities:
\begin{equation}
 -\dot{B_z}=\nabla^2 _h\dot{\mathcal{B}}, 
\label{eq3}
\end{equation}
\begin{equation}
\nabla_h \cdot \dot{\vecB}_h =
\frac{\partial  \dot{B_x}}{\partial x} + 
\frac{\partial \dot{B_y}}{\partial y} = \nabla^2_h 
\left(  \frac{\partial \dot \mathcal{B}}{\partial z} \right)
\label{eq41}
\end{equation}
\begin{equation}
- \veczhat \cdot \left( \nabla \times \dot{\vecB}_h \right) = 
-\frac{4 \pi}{c} \dot{J_z} = 
\frac{\partial  \dot{B_x}}{\partial y} - 
\frac{\partial \dot{B_y}}{\partial x}=\nabla^2 _h\dot{\mathcal{J}},
\label{eq4}
\end{equation}
where 
$\nabla^2_h=\frac{\partial^2 }{\partial x^2}
+\frac{\partial^2 }{\partial y^2} $,
and the Amp\`ere's law for vertical current $\dot{J_z}$ has been used in Eq.~(\ref{eq4}). 
Comparing Faraday's law with Eq.~(\ref{eq2}), we find:
\begin{equation}
 - c \nabla  \times {\evec}=\dot{\bvec} 
=\nabla \times \nabla \times \dot{\mathcal{B}} 
{\bf \hat{z}} +\nabla \times \dot{\mathcal{J}} {\bf \hat{z}}.
\label{eq5}
\end{equation}
For future use, we can rewrite this result as:
\begin{equation}
-c \nabla \times \evec = \dot{\bvec} = - \nabla_h^2 \dot{\mathcal{B}} \veczhat
+ \nabla_h {\partial \dot{\mathcal{B}} \over \partial z} +
\nabla \times \dot{\mathcal{J}} {\bf \hat{z}}.\label{eq6}
\end{equation}
Uncurling the Eq.~(\ref{eq5}), we derive the electric field 
${\evec}$ in terms of $\dot{\mathcal{B}}$ and  $\dot{\mathcal{J}}$:
\begin{equation}
 c  {\evec} = c(E_x,E_y,E_z)=
- \nabla \times \dot{\mathcal{B}} {\bf \hat{z}} 
-\dot{\mathcal{J}} {\bf \hat{z}}-\nabla \psi 
\equiv \underbrace {c {\bf E^{P}}}_{\mbox {inductive}}-
\underbrace {\nabla \psi .}_{\mbox {non-inductive}}
\label{eq7}
\end{equation}

Here,  ${\bf E^{P}}$ is the purely {\it inductive} contribution to 
the electric field determined by PTD potentials, $\dot{\mathcal{B}}$ 
and  $\dot{\mathcal{J}}$. Within this article, ${\bf E^{P}}$ will be 
referred to as P-solution, or the PTD electric field, 
where superscript ``P''
stands for the PTD method. 
At this point, two comments can provide some insight into the nature
of ${\bf E^{P}}$: (1) the horizontal components of ${\bf E^{P}}$ only
depend upon $\dot{\mathcal{B}}$, which was derived from $\dot{B_z}$;
and (2) the vertical component of ${\bf E^{P}}$ only depends upon
$\dot{\mathcal{J}}$, which is derived from $\dot{J_z}$.
The component of the total electric field arising from
$-\nabla \psi$ is the {\it non-inductive} 
contribution, for which the PTD solution to Faraday's law, Eq. (\ref{eq6}), 
reveals no information.  For the simulation data analyzed by 
\cite{Welsch2007}, \cite{Fisher2010} demonstrated that when 
$(-\nabla \psi=0)$, $i.e.$
when non-inductive contributions to the electric field are ignored,
$\vecE^P$ does a poor job of representing the actual electric field.  
The key to a more accurate reconstruction of the electric field is
incorporating other physical constraints, and/or additional observational
information into the solutions for $\nabla \psi$.  Procedures for doing this
will be described further below.  But we first complete our
discussion for obtaining the full $\vecE^P$ (PTD) solution.

To find $\dot{\mathcal{B}}$ 
and  $\dot{\mathcal{J}}$ and hence the $\vecE^P$ solution, we must solve the three Poisson 
equations~(\ref{eq3})-(\ref{eq4}) with well-posed boundary conditions. 
To permit a net gradient in each potential over the field of view, we choose Neumann boundary conditions
\begin{equation}
 \frac{\partial \dot \mathcal{B}}{\partial n}=0
\label{eq42}
\end{equation}
\begin{equation}
 \frac{\partial }{\partial n}\left(\frac{\partial \dot 
\mathcal{B}}{\partial z}\right)=\dot{B}_n-\frac{\partial 
\dot \mathcal{J}}{\partial s}
\label{eq43}
\end{equation}
\begin{equation}
 \frac{\partial \dot \mathcal{J}}{\partial n}=
-\dot{B}_s+ \frac{\partial }{\partial s}\left(\frac{\partial 
\dot \mathcal{B}}{\partial z}\right),
\label{eq44}
\end{equation}
where subscript $n$ denotes components or derivatives in the direction
of the outward normal to magnetogram's boundary, and subscript $s$
denotes components or derivatives in the counter-clockwise direction
along the magnetogram boundary. The first boundary condition,
Eq.~(\ref{eq42}), implies that the tangential component of the
electric field around the magnetogram boundary vanishes, implying that
the average value of $\dot B_z$ within the magnetogram is zero. If the
average value of $\dot B_z$ is not zero (i.e., a change in the flux
balance, or monopole term), then we can add a correction term to the
electric field post-facto (see Appendix C of \cite{Fisher2010}).  The
second and third boundary conditions,
Eq.~(\ref{eq43})-(\ref{eq44}), derived from evaluating Eq.~(\ref{eq6}) at the magnetogram boundary, are degenerate, i.e. there is
a family of coupled non-zero solutions, $ \dot \mathcal{J}$ and
$\left(\frac{\partial \dot \mathcal{B}}{\partial z}\right)$, which
correspond to zero time derivative of the
horizontal magnetic field: 
\begin{equation}
 \frac{\partial }{\partial x}
\left(\frac{\partial \dot \mathcal{B}}{\partial z}\right)=
-\frac{\partial \dot \mathcal{J}}{\partial y}
\label{eq45}
\end{equation}
and
\begin{equation}
\frac{\partial }{\partial y}
\left(\frac{\partial \dot \mathcal{B}}{\partial z}\right)=
\frac{\partial \dot \mathcal{J}}{\partial x}.
\label{eq46}
\end{equation}
Since the solutions to these Cauchy-Riemann equations
each satisfy the two-dimensional Laplace equation, they can be added
to solutions of equations~(\ref{eq41}) and (\ref{eq4}) without
changing the time derivative of the horizontal field on the boundary.
This means that there is some freedom to specify the solutions of
$\dot \mathcal{J}$ or $\left(\frac{\partial \dot \mathcal{B}}{\partial
  z}\right)$ at the boundary. In practice, this means that one could
choose to set the derivative parallel to the boundary of one of these
two functions to zero, while still obeying the coupled boundary
conditions~(\ref{eq43}) and (\ref{eq44}).  To remove the coupling
between the boundary conditions~(\ref{eq43}) and~(\ref{eq44}), we
choose $\frac{\partial \dot \mathcal{J}}{\partial s}=0$ along the
magnetogram boundary, meaning that $\dot \scrJ$ is assumed uniform
along the magnetogram boundary.  (This implies no change in the net
signed vertical current through the field of view.)  This also
allows us to set the value of $E_z$ for $\vecE^P$ to zero at the
magnetogram boundary by simply subtracting the boundary value from the
entire solution for $\dot \scrJ$. Choosing the condition
$\frac{\partial \dot \mathcal{J}}{\partial s}=0$ also means that the
solution of the Poisson equations must be done in a certain order.
First, the solution for $\left(\frac{\partial \dot
  \mathcal{B}}{\partial z}\right)$ must be obtained, then the solution
for $\dot \mathcal{J}$ can be obtained, using the variation of
$\left(\frac{\partial \dot \mathcal{B}}{\partial z}\right)$ along the
magnetogram boundary to specify the normal derivative of $\dot
\mathcal{J}$.  The solution for $\dot \scrB$ is independent of the
other two solutions, and so can be obtained either before or after
obtaining the other two solutions.

When solving Poisson equations~(\ref{eq3})-(\ref{eq4}) and
boundary conditions (\ref{eq42}-\ref{eq44}) numerically, it is 
important to implement a fast and robust numerical scheme, especially 
when dealing with large magnetogram datasets.
In the past, \cite{Fisher2012, Fisher2010} used the 
Newton-Krylov technique adapted 
from the  \texttt{RADMHD}  code \citep{Abbett2007}. Its main 
disadvantage is slow computational speed and poor robustness when applied
to this particular problem.  In this article, we adopt \texttt{FISHPACK}, a 
fast and robust collection of Fortran 
subprograms developed at NCAR \citep{Schwarztrauber1975}.  
\texttt{FISHPACK} applies the cyclic reduction technique \citep{Sweet1974}
and the standard five-point finite difference approximation for the Laplacian
to solve the two-dimensional 
Helmholz equation in Cartesian or spherical (Appendix~\ref{sphere}) coordinates.
This software was designed
to solve Helmholz equations (the Poisson equation is a special case of 
the Helmholz equation) with Neumann, Dirichlet, or periodic
boundary conditions, and can use either a centered or staggered grid.
To convert our equations into a form that is
compatible with the \texttt{FISHPACK} subroutine \texttt{HWSCRT},
we use centered, 2nd-order accurate finite differences to approximate
first derivatives in $x$ and $y$, and the centered 5-point
expression for the Laplacian of a function $f$ in a Cartesian, 
two dimensional geometry:
\be
{\partial f \over \partial x} |_{i,j} = ( f_{i+1,j}-f_{i-1,j} ) / (2 \Delta x),
\label{xderiv}
\ee
\be
{\partial f \over \partial y} |_{i,j} = ( f_{i,j+1}-f_{i,j-1} ) / (2 \Delta y),
\label{yderiv}
\ee
\be
\nabla_h^2 f = (f_{i+1,j} + f_{i-1,j} + f_{i,j+1} + f_{i,j-1} - 4 f_{i,j} )
/ (\Delta x )^2,
\label{5point}
\ee 
where in the last equation we assume $\Delta x = \Delta y$.
Similar to spatial differences, to evaluate time-derivative of function $ f_{i,j}$ at time $t_k$, $ \dot f_{i,j}(t_k)$, the source terms (the left-hand sides) of Eq.~(\ref{eq3})-(\ref{eq4}), we use the centered finite difference approximation
\be
 \dot f_{i,j}(t_k)=(f_{i,j}(t_{k+1})-f_{i,j} (t_{k-1}))/(2 \Delta t),
 \ee
where $\Delta t$ is the time step between two consecutive frames $\Delta t=t_{k+1}-t_k$. 
Once a solution for $\dot \scrB$ has been obtained, the electric field components $E^P_x$ and $E^P_y$ are computed by using the finite
derivative approximations (Eq.~(\ref{xderiv}-\ref{yderiv})) in the expression for $- \nabla \times \dot \scrB \veczhat$;
for $E^P_z$ we use the solution $- \dot \scrJ$. We add a single layer of ghost zones around the periphery of each solution domain
to ensure that the Neumann boundary conditions are obeyed when the above finite difference expressions are used. Given the known Neumann (normal first derivative) boundary conditions (Eq.~(\ref{eq42}-\ref{eq44})) and the interior solution points that are returned from \texttt{HWSCRT}, using the expressions in Eq.~(\ref{xderiv}-\ref{yderiv}), we then determine the values of the solution variables in the ghost zone layers. This procedure leaves the four corners of the ghost zone values undefined; 
for cosmetic reasons when displaying the solution, we assume each corner value is the average of its
two nearest-neighbor ghost zone values.  

As \cite{Schwarztrauber1975} points out, it is
possible to specify Neumann boundary conditions that are incompatible with some Poisson
equations.  Their approach to this quandary is to add a constant value to the source term
of the Poisson equation, with that value varied until the best possible fit to
the boundary values can be obtained.  This offset value is returned by \texttt{HWSCRT} subroutine as the
\texttt{PERTRB} variable, which we record along with the solution itself.  The value of
\texttt{PERTRB} is then monitored to ensure that it remains small compared to typical values
of the source term of our Poisson equations.

Finite difference approximations for the first derivative (Eq.~(\ref{xderiv}-\ref{yderiv}))
and the Laplacian (Eq.~(\ref{5point})) introduce incompatibilities, since the $5$-point Laplacian implicitly assumes that first derivatives are evaluated at half-integer grid locations. These incompatibilities could be eliminated by using a staggered
grid, in which electric field variables are defined at cell edges, while magnetic field
values are defined at cell centers. However, we have found that the resulting 
errors are usually small,
and the introduction of a staggered grid introduces a number of complications,
when adding the
electric fields derived from the gradients of scalar potentials to the $\vecE^P$ solutions.  Here, we have decided to accept these incompatibilities as a cost of using a centered grid,
where the electric field and magnetic field variables are assumed co-spatial. We may adopt a staggered grid in a future version of the software.

One case where the incompatibility errors may not be small, is
if the solutions are dominated by high-frequency noise near the magnetogram boundaries
because of $e.g.$ poor signal-to-noise ratios in areas of weak magnetic fields.
In this case, we have found it useful to add ``zero-padding'' to the solution domain
for a modest number of zones around the periphery
of the observed magnetograms, in which all magnetic field values and their time derivatives
are set to zero, forcing
the source terms of all of the Poisson equations to be zero within a thin 
ribbon inward of
the computational boundary.  The smoothness of the resulting Poisson equation solutions
ensures that the incompatibilities of the finite difference approximations described
above are not important in the boundary regions, and therefore
that the finite difference expressions for the Neumann boundary conditions are
good approximations for the normal first derivatives.

\subsection{Enforcing the Condition $\vecE \cdot \vecB = 0$ with the potential $\psi^I$}
\label{vecEPI}

One of the properties of the derived inductive PTD electric field $\vecE^P$ (Eq.~(\ref{eq7})) is that for most of the cases and even for ideal MHD case, $c \vecE = -\vecv \times \vecB$, that has $\evec$  perpendicular to $\vecB$, the derived $\vecE^P$ is not perpendicular to $\vecB$. To resolve this inconsistency, in this section we review the ``iterative'' method described in \S 3.2 of \cite{Fisher2010} plus the changes that we have made to it since then. This technique will be used extensively in the discussions of other contributions to the electric field  below.

The goal of the iterative method is to construct the gradient of an electric 
potential $\psi^I$ that obeys the constraint
\be
\nabla \psi^{I} \cdot \vecB = c \vecE^P \cdot \vecB ,
\label{idealpsi}
\ee
for all points within the vector magnetogram
domain.  Once found, $\nabla \psi^I$ can then be subtracted
from $\vecE^P$ to yield an electric field which is normal to
$\vecB$, but which has the same curl as $\vecE^P$.  
In practice, and as described in \S 3.2 of \citealt{Fisher2010}, the
procedure is to find two two-dimensional functions, $\psi^I (x,y)$ and 
$\partial \psi^I / \partial z (x,y)$
over the 2-d domain of the vector magnetogram, such that
\be
\nabla_h \psi^I \cdot \vecbhat_h + b_z \partial \psi^I / \partial z = 
c\vecE^P_h \cdot \vecbhat_h + b_z c E^P_z ,
\label{iterative}
\ee
where $\vecbhat_h$ and $b_z$ are respectively
the horizontal and vertical components of the unit vector
$\vecbhat$ that points in the direction of $\vecB$.  Once $\psi^I$ and
$\partial \psi^I / \partial z$ have been found, then we define the quantity
\be
c\vecE^{PI} = c\vecE^P_h - \nabla_h \psi^I +  \veczhat ( c E^P_z -\partial \psi^I / \partial z),
\ee
where $\vecE^{PI}$ then obeys both Faraday's law and the ideal constraint 
$\vecE^{PI} \cdot \vecB = 0$.

We continue to follow the iterative method 
procedures described in \S 3.2 of \cite{Fisher2010}
to determine $\psi^I$ and $\partial \psi^I / \partial z$, but with the following changes:
First, in Step 4 of that procedure, in which a horizontal Poisson equation is 
solved to update the iterative guess for $\psi^I$, we now use the \texttt{\texttt{FISHPACK}} routine
\texttt{HWSCRT} to solve the Poisson equation, instead of using FFT techniques that
implicitly assume periodic boundary conditions.  We assume
homogenous Neumann boundary conditions for $\psi^I$, such that 
$\partial \psi^I / \partial n = 0$ on the outer boundary, meaning that the normal
component of the horizontal electric field due to the scalar potential is zero on 
the outer boundary of the vector magnetogram.  We use equations (\ref{xderiv}-\ref{yderiv})
to approximate the components of the horizontal gradient of $\psi^I$, and equation
(\ref{5point}) to approximate the Laplacian.  Ghost zone values are defined in a similar
fashion as in the previous section for determining $\dot \scrB$.  The solution for 
$\partial \psi^I / \partial z$, in Step 5 of the iterative procedure,
is usually negligibly small at the outer boundary of the magnetogram, provided 
that the horizontal components $\nabla_h \psi^I$ are also small there.
This results in the desired property that $E^{PI}_z$ is essentially zero 
at the outer magnetogram boundary.
An additional change we have made from \cite{Fisher2010} is that instead of evaluating
an error criterion to end the iteration sequence, we have found it more efficient 
to simply specify a maximum number of iterations, typically chosen to be 25
(see discussion in \S \ref{anmhd_intro}).

An important point about the iterative method is that it can be applied not only to the
PTD solution $\vecE^P$ itself, but to any solution for the electric field 
$\vecE$ which contains
components parallel to $\vecB$, that one might like to minimize.  In particular, we can
apply the iterative method to solutions, which include both the PTD solution and additional
electric field contributions from scalar potentials determined from Doppler measurements
or correlation tracking results, as are described in further detail below.  
Once these interim solutions are obtained,
the iterative method can then be applied as a final step to generate the scalar potential 
$\psi^I$ (and $\partial \psi^I / \partial z$) needed to satisfy $\vecE \cdot \vecB=0$.

Finally, we comment on the uniqueness of the solutions for $\psi^I$ found from the iterative
method, given a fixed input electric field $\vecE$.
As noted in \cite{Fisher2010}, mathematically the constraint 
(\ref{idealpsi}) does not result in a unique solution for $\psi^I$ and
$\partial \psi^I / \partial z$.  Empirically, however, we find that the iterative procedure of \S 3.2 of \cite{Fisher2010} (and also with the modifications described here) consistently results in the same solution for $\psi^I$ and $\partial \psi^I / \partial z$, even when initialized with substantially different trial functions.  We surmise that applying the iterative procedure itself imposes some additional constraints beyond (\ref{idealpsi}) which we do not fully understand, but which then result in a unique solution $\psi^I$ for the input $\vecE$.  In fact, however, given any $\vecE$ that is consistent with both Faraday's and Ohm's laws, an additional potential functions can be constructed that can be superposed without violating either constraint. So, while our solution method returns a unique answer, Faraday's and Ohm's laws do not, by themselves, fully constrain $\vecE$.

\subsection{Non-inductive Doppler and Fourier Local Correlation Tracking (FLCT) electric fields: 
${\bf E^D}$ and ${\bf E^{FLCT}}$}
\label{pos}

When examining the first and second rows of Figure 2 
of \cite{Fisher2012}, comparing
the actual components of the test simulation electric field 
with the same
components of $\vecE^P$, as well as the actual and inverted vertical
Poynting flux, one of the most obvious discrepancies is the far lower range of
the Poynting flux values originating from the electric field inversion as compared to 
the true
values (see Figure 3 in \cite{Fisher2012}).  The $x$ and $y$ components of $\vecE$ also show significant
discrepancies,  with the derived horizontal components of $\vecE$ being smaller in magnitude than 
the actual values. Is there some additional information about the horizontal electric field 
that can be determined from other observational data besides the 
magnetic field?

Estimating the transport of magnetic energy across the photosphere requires accurately reconstructing photospheric electric fields during flux emergence.  From studying \texttt{ANMHD} simulations of an emerging bipolar magnetic configuration, we have found that, by itself, Faraday's law applied to $\dot B_z$ does not completely capture significant aspects of horizontal electric fields present during magnetic flux emergence. Here we review two possible sources of electric fields
which may not be reflected in solutions of Faraday's law:  Electric
fields determined from Doppler measurements, and those determined using
correlation tracking methods.  In the discussion below, we will first
consider the computation of these electric fields from the conceptually
easier vantage point of zero viewing angle, where line-of-sight is
normal to the solar surface (\S \ref{DEFzero} and \S \ref{FLCTzero}).  Once we have introduced all of the physical concepts, we will describe the modifications necessary to account for
non-normal viewing angles (\S \ref{nonzero}).

\subsubsection{Doppler Electric Fields: Normal Viewing Angle}
\label{DEFzero}

Certainly in flux-emergence
regions where the vertical magnetic field is near zero ($i.e.$ near PILs), 
a knowledge of the vertical velocity $V_z$ would yield an unambiguous 
determination of the horizontal electric field:  
\be
c \vecE^{PIL}= -V_z \veczhat \times \vecB_h,
\label{epil}
\ee
if we can assume that an ideal MHD Ohm's law applies in the photosphere.
Such measurements could be determined by Doppler velocity measurements viewed
from directly above the active region. Figure 1 of \cite{Fisher2012} illustrates with a thought-experiment
that such a flux-emergence electric field might even have zero curl near
the PIL, and yet, if the Doppler velocity component $V_z$ had a large amplitude along the PIL region, would
still result in a very large electric field. One possible approach to incorporating such data would
be to simply replace the PTD solution $\vecE^P$ with $\vecE^{PIL}$ for
those parts of the magnetogram near the PIL location.  If one did this,
however, the resulting solution near the PIL
would not necessarily obey Faraday's law, 
and furthermore, could vary discontinuously and/or unphysically, as one
crossed the boundary between regions where the two solution methods for $\vecE$
were applied. 

The solution to this dilemma is to combine the two solutions in such a way
that Faraday's law is obeyed everywhere in the magnetogram region, but that
the non-inductive (curl-free) component of $\vecE^{PIL}$ is added to
$\vecE^P$, to provide a more realistic electric field estimate in
flux emergence regions.  To do this, we need two additional ingredients:
(1) an estimate for $\vecE^{PIL}$ that one can apply when one 
leaves the vicinity
of the PIL, and (2) a technique for removing the inductive contribution from
$\vecE^{PIL}$, since $\vecE^P$ will contain all the needed
information about Faraday's law.  

Addressing the first of these ingredients, as one moves away from the PIL, 
Eq.~(\ref{epil}) becomes increasingly
uncertain for two reasons:  first, as the vertical component of the field
becomes nonzero, it becomes increasingly possible for any existing
flows parallel to the magnetic field to contribute to the Doppler signal without contributing to
an electric field.  Second, once $B_z$ is no longer zero, there is a
contribution to the horizontal
electric field from horizontal flows crossed into
$B_z \veczhat$, which is not reflected in Eq.~(\ref{epil}).  We adopt
the ad-hoc solution of \cite{Fisher2012}, namely that $\vecE^{PIL}$ is
multiplied by an attenuation or confidence factor, $w$, that reduces 
the amplitude of $\vecE^{PIL}$ as the ratio $|B_z|/|B_h|$ increases 
above zero.  We adopt the same functional form of $w$ introduced in
\cite{Fisher2012}, namely
\be
w=\exp \left[-\frac{1}{\sigma_{PIL}^2} \left| \frac{B_z}{B_h}\right|^2\right] ,
\label{wdefinition}
\ee
where $\sigma_{PIL}$ is an adjustable parameter that can be tuned by comparing
with test case data, where the electric field is known.  The determination of the optimal value of $\sigma_{PIL}$ for the \texttt{ANMHD} test case is discussed in \S \ref{anmhd_intro}.

Addressing the second ingredient, we must find a way of introducing
only the curl-free contribution from the electric field estimate
$w \vecE^{PIL}$.  This can be accomplished if we first write
the resulting electric field in terms of a scalar potential, $\psi^D$,
such that 
\be
c \vecE_h^D = - \nabla_h \psi^D ,
\label{EDdef}
\ee
and then set the divergence of $c \vecE^D$ to the divergence of 
$c w \vecE^{PIL}$, resulting in the Poisson equation
\be
\nabla_h^2 \psi^D = - c \nabla_h \cdot ( w\ \vecE^{PIL}) ,
\label{psiddef}
\ee 
that we solve for $\psi^D$ using \texttt{FISHPACK}. Since
$\vecE_h^D$ is the gradient of a scalar function, it automatically has
zero curl. 

When $\vecE_h^D$ derived from $\psi^D$ is added to the PTD solution $\vecE^P$, we denote the resulting
solution as the PD electric field.  If we then add the
ideal correction $\vecE^I$ to the PD solution, by applying the $\vecE \cdot \vecB=0$ constraint, we denote that resulting electric field as the PDI solution.  

\subsubsection{FLCT-derived Electric Fields}
\label{FLCTzero}

Local Correlation Tracking (LCT) techniques have been used extensively
in recent years to estimate flows in the
solar atmosphere, as noted in the introduction.  
These methods use observed pattern motions of some quantity,
such as the specific intensity, or the normal component of the
magnetic field, to provide
an estimate of the velocity components of the plasma
that are parallel to the plane of the image.  The input to LCT
techniques is the following: two images of the quantity that is being tracked,
the time difference between the two images, and an a-priori assumption
about the coherence length of the flow-field.  The output from LCT is a two-
dimensional flow field that warps the first image into the second image
over the elapsed time between the images.  The fundamental assumption of
LCT techniques is
that the observed pattern motion corresponds to real physical motion.
In the remainder of this paper, we will focus on the
use of the Fourier Local Correlation Tracking (FLCT)
 technique \citep{Welsch2004,Fisher2008}, but any other
optical flow method \citep{Schuck2006} 
could be used instead.

When the quantity being tracked is the normal 
component of the magnetic field,
velocity estimates obtained from LCT methods have been
found to provide a good estimate on their own for the ``inductive'' flows
responsible for changes in the normal field (see $e.g.$ Figure 6 of
\citealt{Welsch2007}).   However, LCT results 
also contain information about flows that
do not contribute directly
to the observed changes in the magnetic field tracer, and
thus provide additional information about the velocity field 
(or the electric field, assuming an ideal Ohm's law).
The ILCT method \citep{Welsch2004} uses this knowledge to explicitly separate 
the flow field into inductive and non-inductive contributions.

In our case, $\vecE^P$ already contains the ``inductive'' 
contribution to the electric
field, so a straightforward addition of the electric field from
flows derived from FLCT would add
redundant information that has already been incorporated, 
while also adding new information from the
non-inductive flows.  We therefore want to determine only the non-inductive
contribution to the electric field from FLCT-derived flows, and add that
result to $\vecE^P$.  The technique for doing this is very similar to how we
add the non-inductive electric field determined from the Doppler measurements. 

Assuming the ideal Ohm's law, we can write the electric field 
derived from FLCT as
\be
c \vecE = - (\vecv_h \times B_z \veczhat) - (\vecv_h \times \vecB_h),
\label{EFLCTRAW}
\ee
where $\vecv_h$ here is the horizontal flow-field from FLCT.  
This results in a contribution to both $\vecE_h$ (the first term in Eq.~(\ref{EFLCTRAW})), and $E_z$, the second term in that equation.
Considering first the contribution of FLCT to $\vecE_h$,
we note that as one approaches the PIL, this estimate for $\vecE_h$
becomes increasingly suspect since $B_z$, the quantity being tracked,
becomes very small, and the validity of the correlation tracking paradigm
to describe the evolution of $B_z$ becomes increasingly questionable.  Since
the FLCT results get unreliable precisely where we
think the Doppler electric field estimate
becomes reliable, we make the Ansatz that
the FLCT-derived electric field should be multiplied by the complement 
($1-w$) of
the same confidence factor, $w$, used to modulate the Doppler electric field.

To include only the non-inductive contribution, we write
the electric field in terms of the gradient of a scalar potential,
\be
c \vecE_h^{FLCT} = - \nabla_h \psi^{FLCT} .
\label{EFLCTDEF}
\ee
This electric field has no curl, and hence no inductive contribution.
To determine $\psi^{FLCT}$, we multiply the right hand side of
Eq.~(\ref{EFLCTRAW}) by $(1-w)$, take its horizontal
divergence, and then set that equal to the divergence of Eq.~(\ref{EFLCTDEF}):
\be
\nabla_h^2 \psi^{FLCT} = \nabla_h \cdot (1-w) \vecv_h \times B_z \veczhat .
\label{PSIFLCTEQN}
\ee
The result of this solution is an electric field that includes the
non-inductive contributions from FLCT which can then be added to $\vecE_h^P$.
This procedure is very similar to ILCT \citep{Welsch2004}, apart from (1) using
electric fields instead of flow-fields and (2)
the use of the confidence factor.

Now, we consider the contributions to $E_z$ from the second term in equation
(\ref{EFLCTRAW}), and how to correct this for the inductive contribution 
to $E_z$ from $\vecE^P$.  First, note that the FLCT estimate for 
$E_z$ is approximately correct, but has a lot of scatter (see $e.g.$ 
the second panel of Figure 11 in \citep{Welsch2007}).
Our own comparison of $E_z$ from FLCT with that from $E_z^P$ shows that the
FLCT contribution contains significant information that overlaps with $E_z^P$.
Since the inductive contribution is already completely 
contained within $c E_z^P = -\dot{\mathcal{J}}$, we need to remove any
inductive signature from the FLCT estimate for $E_z$.  The result of doing this
is a residual contribution to $E_z$ which satisfies the horizontal
Laplace equation, but obeys the Dirichlet boundary conditions for $E_z$ 
that are returned from FLCT.  If the boundary electric fields are near zero, 
as we generally desire, then the resulting FLCT residual $E_z$
contribution is generally small, since solutions of Laplace's equation achieve their extremum values at the boundaries.  The residual 
$E_z$ contribution is certainly very small in the validation 
test case considered later in this paper, as well as for our 
planned HMI inversions,
and we therefore do not include it in our inversion estimates at this time.

When the $\vecE_h^{FLCT}$ solution is added to the PTD electric field $\vecE^P$,
we denote the resulting electric field as the PF solution.  If the ideal
electric field is added to that solution to impose the constraint
$\vecE \cdot \vecB=0$, we denote the resulting electric field as the PFI
solution.  If the FLCT electric field and the Doppler electric field are
added to $\vecE^P$, we denote the resulting electric field solution as
the PDF solution.  Finally, if we impose the ideal electric field constraint
to the PDF solution, we obtain what we denote as the PDFI solution.
As discussed in greater detail later in this article, we generally find that
the PDFI solution provides the best results when compared to the electric
fields in our standard test case.

\subsubsection{Doppler Electric Fields: Non-normal Viewing Angles}
\label{nonzero}

Viewing the solar surface from a non-normal angle ( $\theta \neq 0$) introduces some
complications for estimating the electric field.  These are due to (1)
foreshortening effects, and (2) changes in the direction of the flows
derived from Doppler measurements.  Because our assumed magnetic field
data is from vector magnetograms, we can: correct it for
foreshortening; interpolate it to a uniform grid on the solar surface;
and reproject the magnetic field directions into the locally normal
and horizontal components.  This means that the PTD inversion
technique for $\vecE^P$ can be applied to the remapped data without
any change in the method, as can FLCT, to derive part of the
non-inductive the electric field.  Using the Doppler velocity to
compute non-inductive electric fields along the solar surface,
however, becomes more difficult, and some additional steps must be
taken.

\begin{figure*}[htb]
\centering
\resizebox{0.5\hsize}{!}{\includegraphics[angle=0]
{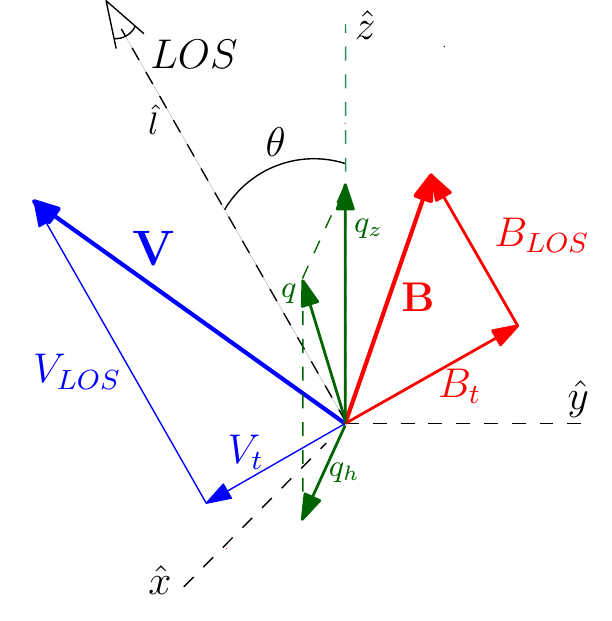}}
\caption{Illustration of the relationship between the LOS direction and 
  the transverse and LOS components of $\vecB$ and $\vecv$ and $\vecqhat$.}
\label{eyeball}
\end{figure*}

We first note that if the angle of the observation is written in terms
of the unit line-of-sight (LOS) vector $\veclhat=[\sin(\theta),0,\cos(\theta)]$, where the direction of this vector is toward
the observer as measured from the solar surface, then the velocity vector
defined by the Doppler shift is given by 
\be
\vlos = ( \vecv \cdot \veclhat ) \veclhat ,
\label{vlosdef}
\ee
where $\vecv$ is the plasma velocity at the source point on the solar surface.  The geometry is shown in Figure
\ref{eyeball}.

The electric field due to the Doppler velocity $\vlos$ near LOS PILs
is then given by
\be
c \vecE = -{\bf V_{LOS}} \times \vecB_t ,
\label{ELOSraw}
\ee
where here $\vecB_t$ represents the components of $\vecB$ in the directions
transverse to the LOS direction.  Note that this electric
field is no longer confined to directions parallel to the solar surface,
but instead is confined to the directions normal to $\veclhat$.
Similar to the normal-viewing angle case, we anticipate that as the ratio
$|B_{LOS}| / |B_t |$ increases above zero, our confidence in this
electric field to represent the transverse electric field decreases due to
possible parallel flows and other contributions to the transverse electric
field. Therefore we multiply $\evec$ by an ad-hoc confidence factor
\be
w_{LOS}=\exp\left[-\frac{1}{\sigma_{PIL}^2} \left| \frac{B_{LOS}}{B_t}\right|^2\right] ,
\label{wlosdef}
\ee
where $\sigma_{PIL}$ has the same value as $\sigma_{PIL}$ in Eq.~(\ref{wdefinition}) for $w$.  Note 
that the confidence factor used for 
determining $E^{FLCT}$ is $1-w$ and not $1-w_{LOS}$, since the FLCT formalism
is not affected by non-normal viewing angles.

To determine non-inductive contributions from the Doppler electric field, we
borrow the iterative technique from \S \ref{vecEPI}, and
find a scalar potential function $\psi^D$ that obeys the constraint
\be
\nabla \psi^{D} \cdot \vecqhat =  w_{LOS} \vlos \times 
\vecB_t \cdot \vecqhat ,
\label{psidopplernonzero}
\ee
where $\vecqhat$ is the unit vector pointing in the direction of
$\vlos \times \vecB_t$.  Note the analogy between the roles of $\psi^D$
and $\vecqhat$ in Eq.~(\ref{psidopplernonzero}) and $\psi^I$ and
$\vecbhat$ in Eq.~(\ref{idealpsi}). Accordingly, as in Eq.~(\ref{iterative}), we then decompose Eq.~(\ref{psidopplernonzero})
into the directions parallel and perpendicular to the solar surface:
\be
\nabla_h \psi^D \cdot \vecqhat_h + q_z \partial \psi^D / \partial z =
w_{LOS} ( \vlos \times \vecB_t )_h \cdot \vecqhat_h + q_z
w_{LOS} ( \vlos \times \vecB_t )_z ,
\label{psid-horizontal-vertical}
\ee
where $\vecqhat_h$ and $q_z$ represent the horizontal and vertical components
of $\vecqhat$, respectively.  We then solve Eq. ~(\ref{psid-horizontal-vertical}) for $\psi^D$ and $\partial \psi^D / \partial z$,
using exactly the same ``iterative'' procedure described in \S \ref{vecEPI}.
The result is a curl-free electric field which can then be added to $\vecE^P$. Henceforth, even if the viewing angle is normal to the solar surface, we use the non-normal formalism for computing electric fields from Doppler shifts, but then set $\theta=0$ in expression for $\hat l$.

\subsubsection{Combinations of Electric Field Contributions:  A Summary}
\label{esummary}
Now that we have introduced the different possible
electric field contributions, we summarize
the naming convention we use henceforth in this article,
depending on which contributions are included in an electric-field estimate (see Table~\ref{table:esummary}). 
To distinguish between different versions of electric field inversions we 
use combinations of the following four letters: ``P'' (for PTD, inductive), 
``D'' (for Doppler, non-inductive), ``F'' (for FLCT, non-inductive), 
and ``I'' (for ideal, non-inductive).  Solutions 
with an ``I'' in their name have
the ideal-MHD constraint, $i.e.$ the addition of the field from the potential 
$-\nabla \psi^I$, imposed as the last step. We also consider the full FLCT results, denoted FI, the full Doppler electric field results, denoted DI, and their sum, denoted DFI.  For FI, DI and DFI we did not impose $\nabla \psi^{I}$, because their total electric field is already ideal. Consequently, we have a notation for the electric field, helicity and Poynting fluxes computed with a variety of different electric field estimation techniques.

\begin{deluxetable}{llll}
\small
\tablecaption{Summary of Electric Field Notation}
\tablewidth{0pt}
\tablehead{
\colhead{ Name}  
& \colhead{Denoted} 
& \colhead{Equation for $\vecE$}  
& \colhead{Input Data / Constraints}
}
\startdata
P solution\tablenotemark{a} & ${\bf E^{P}}$ & $c\evec=c{\bf E^{P}}$ 
& $\dot B_z$, $\dot J_z$ (generally, ${\bf E}\cdot{\bf B}\neq 0$) \\
PI solution\tablenotemark{b} & ${\bf E^{PI}}$ & $c\evec=c{\bf E^{P}}-\nabla \psi^I$
& $\dot B_z$, $\dot J_z$, ${\bf E}\cdot{\bf B}= 0$ \\
PFI solution\tablenotemark{c} & ${\bf E^{PFI}}$ & 
$c \evec=c{\bf E^{P}}+ c{\bf E_t^{FLCT}}-\nabla \psi^{I}$
& $\dot B_z$, $\dot J_z$, ${\bf E}\cdot{\bf B}= 0$, FLCT output \\
PDI solution\tablenotemark{d} & ${\bf E^{PDI}}$ & 
$c\evec=c{\bf E^{P}}+ c{\bf E^{D}}-\nabla \psi^{I}$ 
& $\dot B_z$, $\dot J_z$, ${\bf E}\cdot{\bf B}= 0$, Dopp. data \\
PDFI solution & ${\bf E^{PDFI}}$ & 
$c\evec=c{\bf E^{P}}+ c{\bf E^{D}}+c{\bf E_t^{FLCT}}-\nabla \psi^{I}$
& $\dot B_z$, $\dot J_z$, ${\bf E}\cdot{\bf B}= 0$, Dopp. \& FLCT \\ \hline
FI solution & ${\bf E^{FI}}$ & 
$c{\bf E}=-{\bf V_h} \times {\bf B}$\tablenotemark{e}  
& FLCT data, $V_{LOS}=0$ \\
DI solution & ${\bf E^{DI}}$ & 
$c{\bf E}=-{\bf V_{LOS}} \times {\bf B}$\tablenotemark{e}  
& Dopp. data, $V_h=0$ \\
DFI solution & ${\bf E^{DFI}}$ & 
$c{\bf E}=-{\bf V} \times {\bf B}$\tablenotemark{e} 
& Dopp. \& FLCT data\\
\enddata
\tablenotetext{a}{P for PTD, $^{b}$I for ideal, $^{c}$F for FLCT, $^{d}$D for Doppler, $^{e}$For FI, DI and DFI we do not include $\nabla \psi^{I}$ because the total electric field is already ideal.} 
\label{table:esummary}
\normalsize 
\end{deluxetable}

The biggest changes from the solutions described here and those of 
\cite{Fisher2012} are the adoption of the \texttt{\texttt{FISHPACK}} software to 
solve the two-dimensional Poisson equations, the ability to compute 
contributions to Doppler-shift electric fields from non-normal 
viewing angles, and a much more systematic and quantitative testing of the
accuracy and robustness of the technique (described further in \S \ref{anmhd}).

\section{USING ELECTRIC FIELDS TO COMPUTE POYNTING AND HELICITY FLUXES}
\label{poynting-hel}

\subsection{Poynting Flux Components: {\spot} and {\sf} }
\label{poynting}

The Poynting-flux vector 
\begin{equation}
\label{eq_s}
{\bf S}=\frac{c}{4\pi}({\bf E}\times {\bf B}),
\end{equation}
measures the flow of electromagnetic energy at the photosphere, where 
magnetic field vector ${\bvec}$ is defined from the observations, and 
the electric field vector ${\evec}$ is taken from the techniques described
in \S \ref{ptdtheory}. Since we are
interested in the amount of energy flowing into and out of the corona, 
we focus most of our attention on the vertical component of Poynting 
flux, $S_z$.  This depends upon the horizontal components of both the
electric field and the magnetic field:
\be
S_z = \frac{c}{4\pi}\left(E_x B_y-E_y B_x\right)=
\frac{c}{4\pi}(\vecE_h \times \vecB_h)\cdot \veczhat.
\label{szdef}
\ee Computing $S_z$ is straightforward, given our techniques for
finding all three components of $\vecE$ as described above, and the
availability of the components of $\vecB$ from the vector magnetogram
data.  But we can go further, and decompose $S_z$ into two
contributions, the flux of potential-field energy, and the flux of
free magnetic energy.  The distinction between these two contributions
was first made by \cite{Welsch2006}, who expressed the Poynting flux
in terms of velocities.  An expression for the flux of free magnetic
energy was given in \cite{Fisher2010} in terms of the electric field,
and a decomposition of the horizontal magnetic field in terms of the
poloidal and toroidal potentials.  The basic idea is that the
horizontal magnetic field $\vecB_h$ can be divided into a
potential-field contribution, and a contribution due to currents that
flow into the atmosphere from the photosphere.  Because there are
several ways that one can construct a potential field from the
photospheric vector magnetogram data, we first discuss what we believe
is the best way of performing this decomposition, and then briefly
mention another alternative we have considered.

Horizontal electric fields derived from Faraday's law are derived, in part,
from the observed evolution of the vertical magnetic field.
Therefore, the most self-consistent description of the potential magnetic
field is to find the potential field that best matches the values 
of the measured
normal component of the magnetic field at the photosphere.  
There are a number of ways to do this
using, $e.g.,$ Green's function techniques, or FFTs (though in the latter case,
one must insure the region is flux balanced, and that sufficient padding is
provided outside of the active region to mitigate against artifacts from
periodic boundary conditions).  Once the potential function $\Phi$ has been
derived, the horizontal potential magnetic field at the photosphere
can be evaluated as
\begin{equation}
\vecB_h^P = - \nabla_h \Phi .
\end{equation}
The flux of potential-field energy is then given by
\begin{equation}
S_{z,pot} = \frac{1}{4 \pi} c \vecE_h \times \vecB_h^P .
\label{szpotdef}
\end{equation}
To compute the flux of free magnetic energy, we use the residual horizontal 
field that is the difference between the measured and potential components:
\begin{equation}
S_{z,free} = \frac{1} {4 \pi} c \vecE_h \times \left( \vecB_h - \vecB_h^P \right).
\label{sfreedef}
\end{equation}

We have also experimented with using the potential field formulation
suggested in Appendix A of \cite{Fisher2010}.  This has the advantage
of being extremely easy to compute in terms of the poloidal-toroidal
potential function $\partial \mathcal{B} / \partial z$, which has the
same PTD formalism we use to solve for $\vecE^P$.  However, as
\cite{Welsch2014} have shown, this potential field matches the
observed values of $\nabla_h \cdot \vecB_h$ at the photosphere, and
may not accurately reflect the measured value of $B_z$.  If the actual
magnetic field were in fact potential, \cite{Welsch2014} show that the
two formulations would provide the same result.  Since currents are
present, however, the two formulations can and do show differences in
$B_z$ at the photosphere.  Since our derived electric fields are
driven primarily by changes in $B_z$, we therefore choose a potential
field that explicitly matches $B_z$ at the photosphere.

For short sequences of vector magnetograms, like e.g.  the analysis of the \texttt{ANMHD}
test data described in \S \ref{anmhd}, we compute the potential
magnetic field by using the Green's function technique (see Appendix B of
\citealt{Welsch2014}), as we have found it to be the most robust and accurate.  It is, however, very computationally
intensive, and
can be impractical for large datasets.  
For long cadences of large vector magnetograms, we use a padded FFT
solution for the potential magnetic field, with the size of the needed
padding area calibrated by solutions from the Green's function technique.  


\subsection{Helicity Flux Rate: $\frac{dH_R}{dt}$}
\label{hel-sec}

The {\it magnetic helicity} of magnetic field ${\bf B}$ in a volume $V$ 
\begin{equation}\label{eq_htot}
H = \int {\bf A} \cdot {\bf B} \, dV,
\label{helfluxeqn}
\end{equation}
where 
\begin{equation}\label{eq_htot}
{\bf B}=\nabla \times {\bf A},
\label{vecoit}
\end{equation}
measures the linkage of magnetic field lines and is useful for 
describing their topology. Magnetic helicity is physically meaningful 
only for magnetic fields that are fully contained in a volume $V$, 
a condition which  for active regions  with magnetic field lines extending 
far above and below the photosphere, is not satisfied.  
In this case, subtracting the helicity of the potential 
field ${\bf B_p}$ that matches the normal component of $B_n$ on $S$, 
one can define topologically meaningful and gauge-invariant 
{\it relative magnetic helicity}, i.e. helicity relative to 
the potential field  \citep{Berger1984, Finn1985} 
\begin{equation}\label{eq_htot}
H_R = \int_V ({\bf A}+{\bf A_p})  \cdot ({\bf B}-{\bf B_p})  dV.
\label{finan}
\end{equation}

Due to absence of accurate magnetic field measurements throughout the
corona, it is hard to derive relative magnetic helicity directly from
the observations.  Instead, rate of change of relative helicity, $d
H_R/ dt$ is used to characterize the coronal magnetic field
complexity. If we know the electric field, then the
time-rate-of-change of relative helicity in volume $V$ --- equivalent to the flux of helicity into $V$ --- is given by the surface integral (Eq.~(62) in
\cite{Berger1984})
\begin{equation}
\left(\frac{dH_R}{dt} \right) =
-2 \int \left({\bf A_p}\times {\bf E} \right) \cdot {\bf \hat{z}} \, da=
-2 \int \left(A_{px}E_y-A_{py}E_z\right) da,
\label{eq_hele}
\end{equation}
where ${\bf A_p}=\left(\frac{\partial \mathcal{B}}{\partial y},
-\frac{\partial \mathcal{B}}{\partial x},0\right) = \nabla \times
\mathcal{B} \veczhat$ is the vector potential that generates the
potential field ${\bf B_p}$ in $V$, which matches the photospheric
normal field $B_z$.  We require that ${\bf A_p}$ obey the Coulomb
gauge condition (divergence-free in $V$) and be tangential on the
bounding surface. For a closed magnetic field rooted in
the photosphere, the helicity flow across the upper boundaries is zero
and only photospheric surface term remains.

If we have an ideal electric field
$c{\bf E}=-{\bf V}\times{\bf B}$, 
then the helicity flux rate, given by Eq.~(\ref{eq_hele}), 
splits into two terms \citep{Berger1984_rcc}:
\begin{equation}
\left(\frac{dH_R}{dt} \right) =
  \underbrace{-2\int  ({\bf A_p}\cdot {\bf V_h} ) B_z da}_{\mbox{braiding, FI}}+
  \underbrace{2\int ({\bf A_p}\cdot {\bf B_h} ) V_z da} _{\mbox{emergence, DI}}.
\label{eq_helv}
\end{equation}
The first term corresponds to magnetic helicity generated by flux tubes 
moving horizontally in the photosphere (braiding), and the second 
term corresponds to helicity injection due to  emergence from 
the solar interior into the corona.  Note, that adopting the naming convention from \S~\ref{esummary}, braiding and emergence terms would correspond to the FI and DI electric-field solutions respectively, and their sum to the DFI solution.  When using the electric field form of the relative helicity flux (Eq.~(\ref{eq_hele})), the braiding and
emergence terms are not neatly separated. 

In \S\ref{anmhd}, using \texttt{ANMHD} test data with a known helicity 
flux rate, we use both the electric field (Eq.~(\ref{eq_hele}), PDFI-, PDI-, or PFI-solution) and 
the velocity field (Eq.~(\ref{eq_helv}), FI-, DI-, or DFI-solution) to evaluate the best approach for estimating $dH_R/dt$. 
 

\section{VALIDATION OF THE ELECTRIC FIELD INVERSION METHOD USING {ANMHD} SIMULATIONS}\label{anmhd}
\subsection{ANMHD Run: Input, Output, Parameters}\label{anmhd_intro}

In this paper, we use a specific set of anelastic pseudo-spectral
\texttt{ANMHD} simulations \citep{Fan1999,Abbett2000, Abbett2004} of
an emerging magnetic bipole in a convecting box \citep{Welsch2007} to
test our improved electric field inversion technique. From \texttt{ANMHD} magnetic fields and plasma velocities, we know the actual
electric fields, which we can compare  to electric fields derived using
the simulation's evolving magnetic field.
In the past, this \texttt{ANMHD} simulation has been used for several
studies of velocity field inversions \citep{Welsch2007, Schuck2008},
as well as in test-cases for the first electric field inversions
\citep{Fisher2010, Fisher2012}.  \cite{Fisher2012} showed that the
PDFI-solution significantly improves the accuracy of the derived
PI-solution, and the improvement from the knowledge of the Doppler
velocity (PDI) is significantly more important than that of the
horizontal velocity (PFI), at least in this example of magnetic flux
emergence.
In this paper, we perform a series of improvements to PDFI method
beyond \cite{Fisher2012}: we expand the derivation of the
non-inductive contribution to non-normal viewing angles, introduce
spherical coordinates for the PTD solution (Appendix~\ref{sphere}) and
significantly speed up the Poisson equation solutions by using
\texttt{FISHPACK}. With the above upgrades, we find that the PDFI
method yields a good estimate of the electric field, Poynting and
helicity fluxes and is ready to be applied routinely to the observed
vector magnetograms.
%


We use a pair of \texttt{ANMHD} vector magnetograms, separated by
$\Delta t=250$ s with a pixel size of $\Delta s=348.36$ km, and a LOS
velocity map, observed at a specific viewing angle, to derive the
electric field and the vertical Poynting flux for a given set of
parameters (see Table~\ref{tparam}). To estimate the horizontal
velocity field, $(V_x, V_y)$, we use the Fourier local-correlation
tracking (\texttt{FLCT}) technique \citep{Welsch2004, Fisher2008} (\verb+http://solarmuri.ssl.berkeley.edu/~fisher/public/software/FLCT/C_VERSIONS/+). We
denote the Gaussian window size scale (a parameter in FLCT) by
$\sigma_{FLCT}$. To suppress noisy behaviour, we only calculate
velocities where $|B_z|>370$ G, or $5\%$ of maximum $|B_z|$
\citep{Welsch2007}.

In Table~\ref{tparam} we summarize all the input and output variables
of the PDFI run and their typical ranges. Further, we
vary the observed viewing angle $\theta$ within $[0,60]^{\circ}$-range
to estimate the accuracy of the PDFI method at non-zero viewing
angles. We also vary the other free parameters: the number of iterations in
the perpendicularization technique ($N_{iter}$, see \S~\ref{vecEPI}),
the width of the Gaussian window used in the \texttt{FLCT} estimate for the
horizontal velocity ($\sigma_{FLCT}$, see below) and the width of
polarity inversion line ($\sigma_{PIL}$, see \S~\ref{DEFzero}) to find
the best values of the parameter set that yields the most accurate electric
field solution (Right column).

For the remainder of this paper, to assess the performance of our
methods, i.e. of the reconstruction $u'$ of the \texttt{ANMHD}
variable $u$, we use the following metrics: (1) a fraction of the integrated total 
$\displaystyle f(u,u')=\frac{\sum_i u'}{\sum_i u}$, (2) the slope or
linear coefficient in the least-squares, polynomial fit, $a(u,u')$:
$u' \approx a_0+a(u,u')*u$, (3) the linear Pearson correlation
coefficient $\rho(u,u')=\frac{{\rm cov}(u,u')}{\sigma_u \sigma_{u'}}$,
where $\sigma_u$ is a standard deviation of $u$, and (4) the
normalized error of $u'$, $Err.=\sigma(u'-u)/\sigma_{u}$, $Err.=0.1$
means that the error of the reconstruction is $10\%$ relative to the
characteristic range of $u$.  Note that to exclude the weak-field
background of $u$, where the observed magnetic field (e.g. in HMI/SDO)
tends to be noisy, we estimated $f,a, \rho$ and $err.$ in locations
where $|B|=({B_x}^2+{B_y}^2+{B_z}^2)^{1/2}>370$ G. In the HMI vector magnetogram case, a similar threshold will be determined from estimated errors in the magnetic field values. To summarize, the
ideal reconstruction $u'$ of variable $u$ satisfies: $f(u,u')=1$,
$a(u,u')=1$, $\rho(u,u')=1$ and $Err.=0.0$.

\begin{table*}[h] \caption{Input and output variables of a set of PDFI runs for the \texttt{ANMHD} test case and their typical range. To test the PDFI method we varied the parameters within a shown range. We then used the best-value parameter shown in the parenthesis as a default.} 
\begin{center} \begin{tabular}{ccc}
{\it Input}  & Description & Observed Range \\ \hline
${B_{t,x},B_{t,y},B_{LOS}} (x,y)$ & Magnetic field & $[-6000,6000]$ Gauss \\
${V_{LOS}}(x,y)$ &Doppler velocity  & $[-0.5,0.2]$ km/s \\
$\theta$ & Viewing angle & $[0,60]^{\circ}$\\  \hline
{\it Output} &  & \\ 
${V_{x},V_{y}}(x,y)$ & \texttt{FLCT} velocity field & $[-0.4,0.4]$ km/s \\
${E_{x},E_{y},E_{z}}(x,y)$ & PDFI electric field & $[-1,1]$ V/cm \\
${S_{z}(x,y)}$ & Poynting flux & $[-2,6]\times 10^{10}$ ergs/(s cm$^2$) \\ \hline
{\it Parameters} &  & Range (Best value) \\
$N_{iter}$ & No. of iterations in ${\bf E}\cdot{\bf B}$=0& $[0,50]$ iterations $(25)$ \\
$\sigma_{FLCT}$ & Gaussian window width &$[0,15]$ pixels $(15)$ \\
$\sigma_{PIL}$ &  PIL width & $[0,2]$ pixels $(1)$\\ 
\end{tabular} \label{tparam} \end{center} \end{table*}

 \begin{figure*}[htb]
  \centering 
  \resizebox{1.1\hsize}{!}{\includegraphics[angle=0]{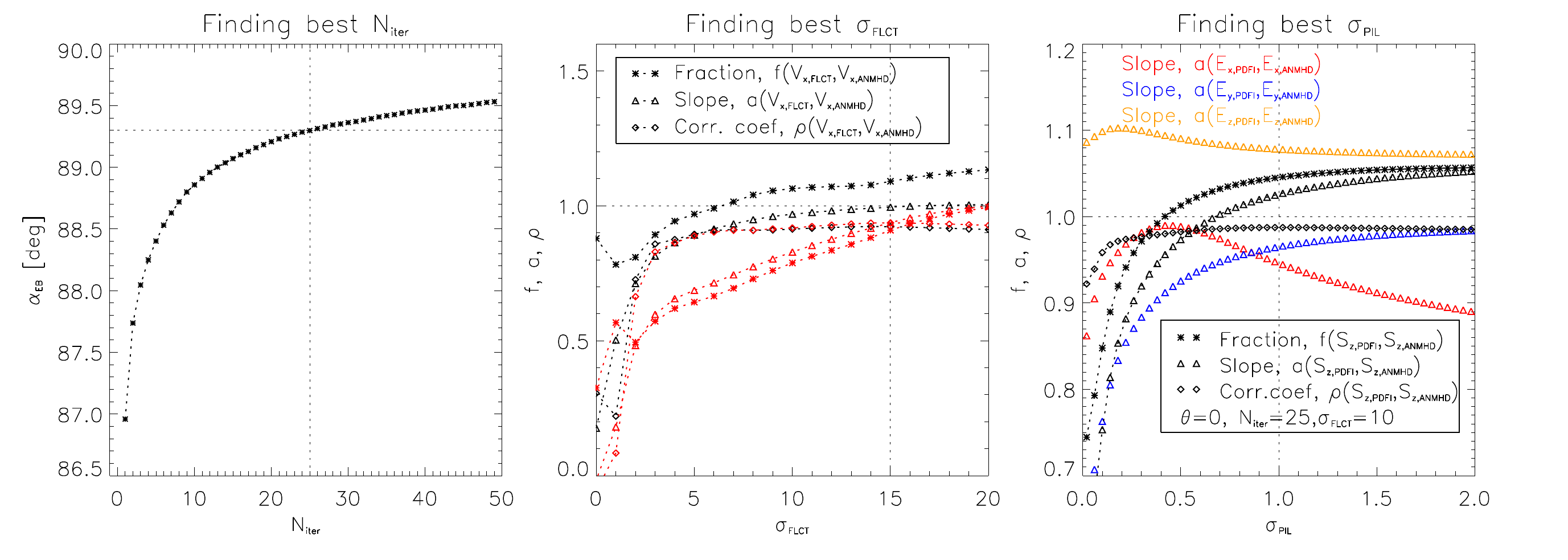}} 
  \caption{Finding best set of parameters for PDFI: $N_{iter}$,
    $\sigma_{FLCT}$ and $\sigma_{PIL}$. {\it Left:} Dependence of the
    angle between E and B on the number of steps in the
    perpendicularization process, $\alpha_{EB}(N_{iter})$. {\it
      Middle:} Quality, $(f,a,\rho)$, of horizontal velocity
    components, $V_x$ (black) and $V_y$ (red), reconstructed using
    \texttt{FLCT} technique at different values of the Gaussian window
    width, $\sigma_{FLCT}$. {\it Right:} Black curves show the quality, $(f,a,\rho)$, of the
    $S_z$ reconstruction at different PIL widths, $\sigma_{PIL}$. Red,
    blue and yellow colors show the slopes 
    $a(E_{x,PDFI},E_{x,ANMHD})$, $a(E_{y,PDFI},E_{y,ANMHD})$ and
    $a(E_{z,PDFI},E_{z,ANMHD})$ respectively, they quantify the faithfulness of the $E_x$, $E_y$ and $E_z$ reconstructions at different values of $\sigma_{PIL}$. Vertical dotted lines show the best values of
    parameters that we further use as default.}
  \label{fig_iter}
\end{figure*}

Figure~\ref{fig_iter} shows three panels that quantitavely justify
selection of the best set of PDFI parameters (vertical dotted lines):
number of iterations ($N_{iter}$, left panel), width of the Gaussian
window ($\sigma_{FLCT}$, middle panel), and the PIL width
($\sigma_{PIL}$, right panel).  The left panel,
$\alpha_{EB}(N_{iter})$, shows dependence of the RMS angle between the
electric and magnetic field vectors on the number of iterations using the
perpendicularization technique (see \S~\ref{vecEPI}).  When looking
for the optimal number of iterations $N_{iter}$, we try to keep
$N_{iter}$ as low as possible to achieve a high-speed performance,
while still aiming for an angle close to $90^\circ$. For the
\texttt{ANMHD} case, without perpendiculatization ($N_{iter}=0$), the
angle between the magnetic and electric field vectors
$\alpha_{EB}=75^{\circ}$. After only one iteration ($N_{iter}=1$),
$\alpha_{EB}=87^{\circ}$. The angle slowly increases to
$\alpha_{EB}=89.3^{\circ}$ by $N_{iter}=25$. The convergence rate slows down, and 
reaches $\alpha_{EB}=89.5^{\circ}$ by $N_{iter}$=50. We chose
$N_{iter}=25$ as the optimal number of iterations, since above $25$
the convergence of $\alpha_{EB}$ toward $90^\circ$ is too slow to justify the additional
computational effort.  We also use $N_{iter}=25$ when applying the non-normal viewing angle technique 
(\S~\ref{nonzero}) for finding the needed scalar potential $\psi^D$.

The Middle Panel of Figure~\ref{fig_iter} shows how the quality of the horizontal velocity reconstruction, using the \texttt{FLCT} technique in the \texttt{ANMHD} test case, depends on the Gaussian window width $\sigma_{FLCT}$. Since velocities parallel to the magnetic field do not affect the time evolution of the magnetic field and hence the fluxes of magnetic energy and helicity, in this plot we only compare components of the flow field that are perpendicular to ${\bf B}$.  We find that  $\sigma_{FLCT}<6$ pixels yield the worst agreement with the actual horizontal \texttt{ANMHD} plasma speed;  $\sigma_{FLCT}=15$ pixels yields  the best agreement, we therefore adopt it as the default value. For comparison, in \cite{Welsch2007} the optimal Gaussian window size was also chosen to be $\sigma_{FLCT}=15$ and in \cite{Fisher2012} $\sigma_{FLCT}=5$.

Finally, the Right Panel of Figure~\ref{fig_iter}, ``Finding the best $\sigma_{PIL}$'', shows the quality of the PDFI electric field components (red, blue, orange) and Poynting flux (black) reconstruction for different  $\sigma_{PIL}$'s, a free parameter that reflects the lack of confidence in the accuracy of the horizontal Doppler electric field away from PILs (see \S~\ref{pos}).  The panel shows that $\sigma_{PIL}$ has a  small effect on the quality of $E_y$, $E_z$ and $S_z$,  above $\sigma_{PIL}=1$.  We find that $\sigma_{PIL}\simeq1$ yields the best ratio between the total reconstructed and \texttt{ANMHD} Poynting fluxes: $\rho\simeq1,a\simeq1$.  

Using the best set of parameters ($N_{iter}=25$, $\sigma_{FLCT}=15$ and $\sigma_{PIL}=1$) found above, in \S~\ref{anmhd_esh},  we estimate the quality of the PDFI reconstructed electric fields, helicity and Poynting fluxes and also estimate the uncertainties in the results at a zero viewing angle. In \S~\ref{anmhd_ang} we describe how observing at the non-zero viewing angles affects the quality of the reconstruction.  Finally in \S~\ref{anmhd_nind} we estimate the roles that different non-inductive Doppler- and FLCT contributions play in the reconstruction and compare current results to \cite{Fisher2012}.

\subsection{Results: Electric Field, Poynting and Helicity Fluxes}\label{anmhd_esh}
 \begin{figure*}[htb]
  \centering 
  \resizebox{0.9\hsize}{!}{\includegraphics[angle=0]{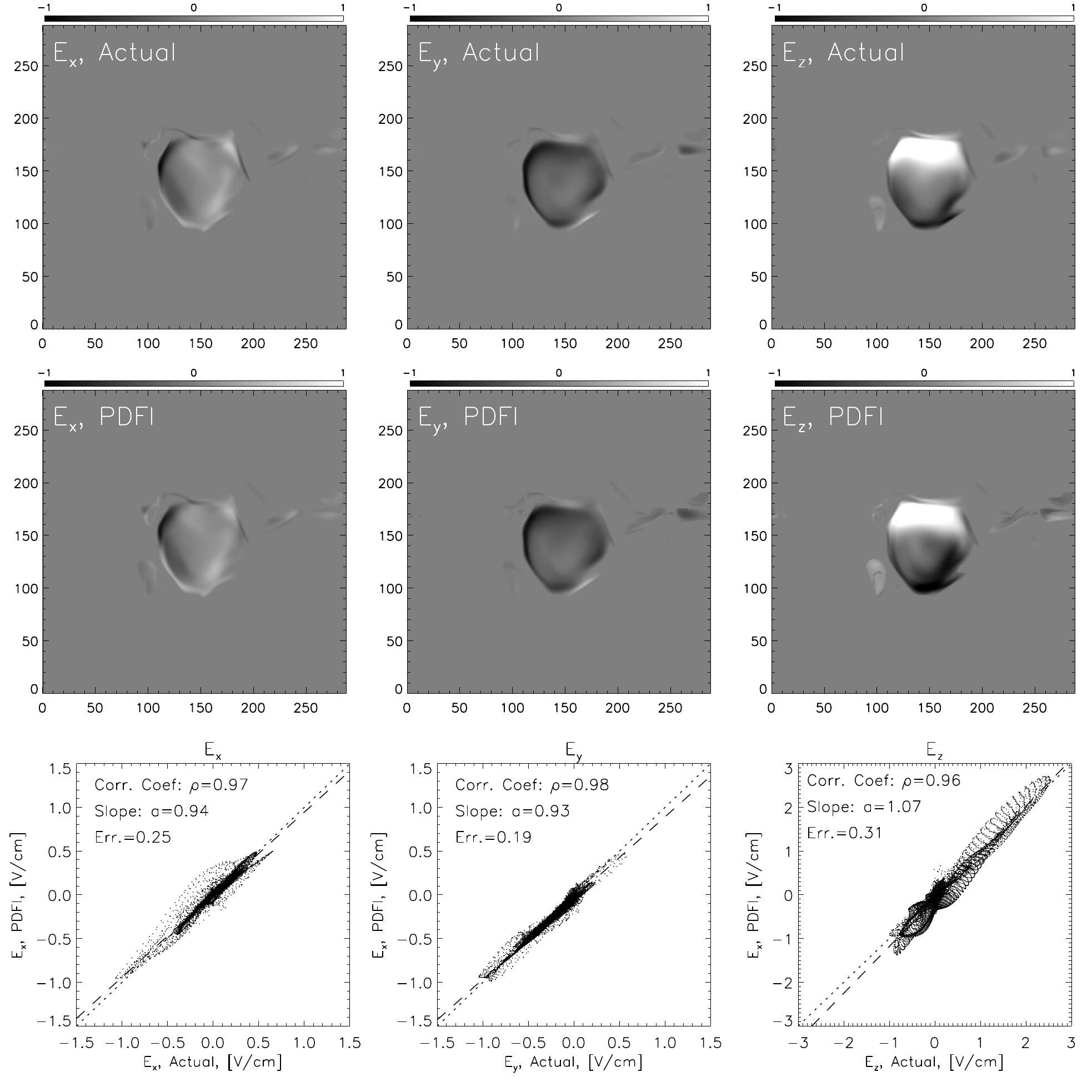}} 
  \caption{Validation of the PDFI electric field at $\theta=0^\circ$:  Actual ({\it top row}) and PDFI ({\it middle row}) electric field vector components, $[E_x, E_y,E_z]$, (left, middle, right) for the \texttt{ANMHD} test-case. {\it Bottom row:} Pixel-by-pixel scatter plots comparing the top two rows. The slopes of the linear fits and correlation coefficients are given in the top left corners.}
  \label{fig_el}
\end{figure*}

   \begin{figure*}[htb]
  \centering 
  \resizebox{0.98\hsize}{!}{\includegraphics[angle=0]{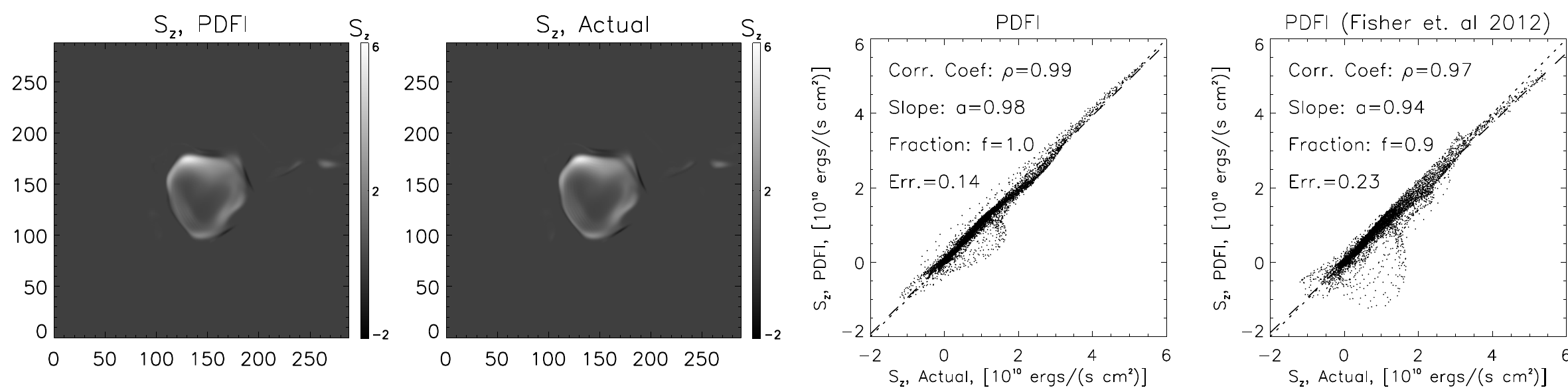}} 
  \caption{Validation of the PDFI Poynting flux at $\theta=0^\circ$. PDFI  ({\it left}) and the actual ({\it middle left}) Poynting fluxes, $S_z$, for the \texttt{ANMHD} test-case, and also the pixel-by-pixel comparison between the two  ({\it middle right}). The {\it far right} panel shows the same comparison, but instead of the current version of the PDFI we use the  \cite{Fisher2012} method. }
  \label{fig_sz0}
\end{figure*}
Figures~\ref{fig_el} and~\ref{fig_sz0} show validation plots that compare electric field components $(E_x,E_y,E_z)$ and vertical Poynting fluxes derived from the PDFI-method with the actual \texttt{ANMHD} quantities. The PDFI-method reconstructs the \texttt{ANMHD} electric-field components quite well: the top and middle rows of Figures~\ref{fig_el} look almost identical. The slope of the linear fit to the reconstructed versus the actual electric-field component ranges from $a=0.94$ to $a=1.07$. The correlation coefficient, describing the quality of linear fit, in all cases is close to one.  Using these electric-field components, we also find a good agreement for the vertical Poynting flux (Figure~\ref{fig_sz0}): the correlation coefficient $\rho=0.99$, the slope $a=0.98$ and the fraction, $f=1$. For comparison, \cite{Fisher2012} found slightly worse results: $\rho=0.97, a=0.94, f=0.9$ (Right Panel).  It should be noted that the MEF method \citep{Longcope2004c} also accurately reconstructed the total Poynting flux in the tests by \cite{Welsch2007}, but the spatial correlation between the true and reconstructed fluxes was significantly worse, at $\rho=0.85$ (see their Figure 14). Decomposing the total Poynting flux into potential and free components (see \S~\ref{poynting}), we find that the PDFI reconstructs $100\%$ of both the potential and free components ($f=1$) and the slope between the reconstructed and PDFI is one ($a=1$). The free and potential components comprise $87\%$ and $13\%$ of the total unsigned Poynting flux respectively. 

We also test how sensitive the Poynting fluxes are to errors in the vertical Doppler velocity. We find that if there is a random Doppler velocity noise of $0.05$-km/s amplitude ($\bar{v}_z=0$), i.e. around $10$ to $20\%$ of the signal, then it does not significantly affect the Poynting flux:  the error increases slightly from $0.14$ to $0.15$, the slope remains close to one ($a=0.98$) and the fraction  $f=1.0$. However, if there is a bias Doppler velocity that increases
all the velocities by $0.05$ km/s (towards the viewer) ($\bar{v}_z=0.05$), then the slope and the fraction increase to $a=1.2$ and the error is $Err.=0.2$. Similarly, if all velocities decrease by $0.05$ km/s ($\bar{v}_z=-0.05$), then the slope decreases to $a=0.8$ and the fraction to $f=0.8$. This test demonstrates how important it is to remove the Doppler velocity bias \citep{Welsch2013}, when inferring electric fields from the observations. 

\begin{figure*}[htb]
  \centering 
  \resizebox{1.0\hsize}{!}{\includegraphics[angle=0]{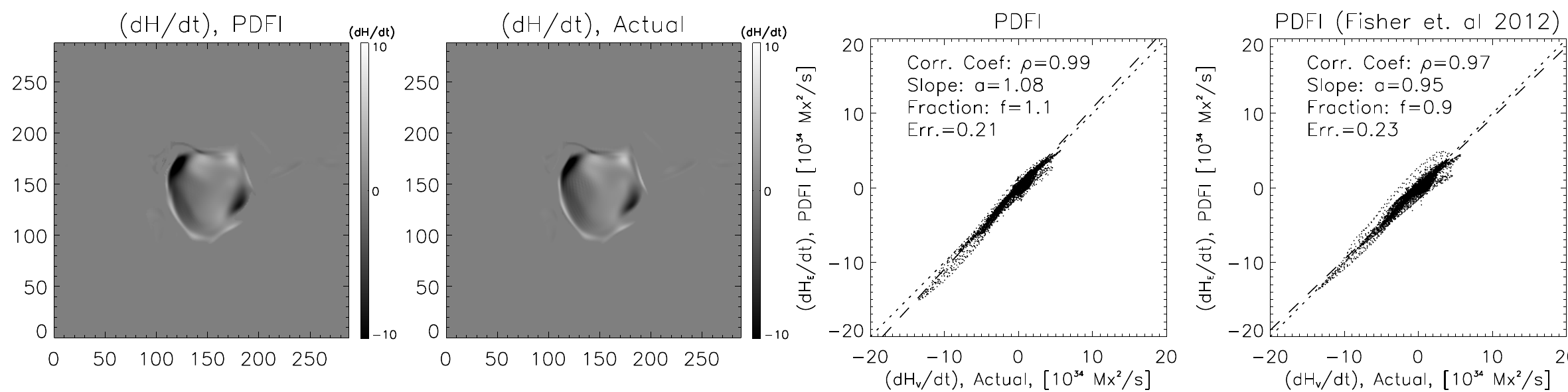}} 
  \caption{Validation of the PDFI helicity flux at $\theta=0^\circ$. See caption of Figure \ref{fig_sz0}.}
  \label{fig_hel0}
\end{figure*}

In Figure~\ref{fig_hel0} we compare actual \texttt{ANMHD} and the PDFI helicity flux rates calculated from $\evec$. We find a good agreement between the two: the correlation coefficient $\rho=0.99$, the slope $a=1.08$, the fraction $f=1.1$ and the error $Err.=0.21$. For comparison, using \cite{Fisher2012} electric fields we find a very similar helicity flux rate, but with a slightly larger scatter ($Err.=0.23$): $\rho=0.97$, $a=0.95$, $f=0.9$.  
The differences between our approach here and that of \cite{Fisher2012} are the adoption of the \texttt{\texttt{FISHPACK}} software to solve the two-dimensional Poisson equations, the ability to compute 
contributions to Doppler-shift electric fields from non-normal 
viewing angles, and a much more systematic and quantitative testing of the
accuracy and robustness of the technique and its parameters.

\subsection{Quality Of Electric Field and Poynting Flux Reconstructions At Non-Zero Viewing Angles}\label{anmhd_ang}

  \begin{figure*}[htb]
  \centering 
  \resizebox{1.0\hsize}{!}{\includegraphics[angle=0]{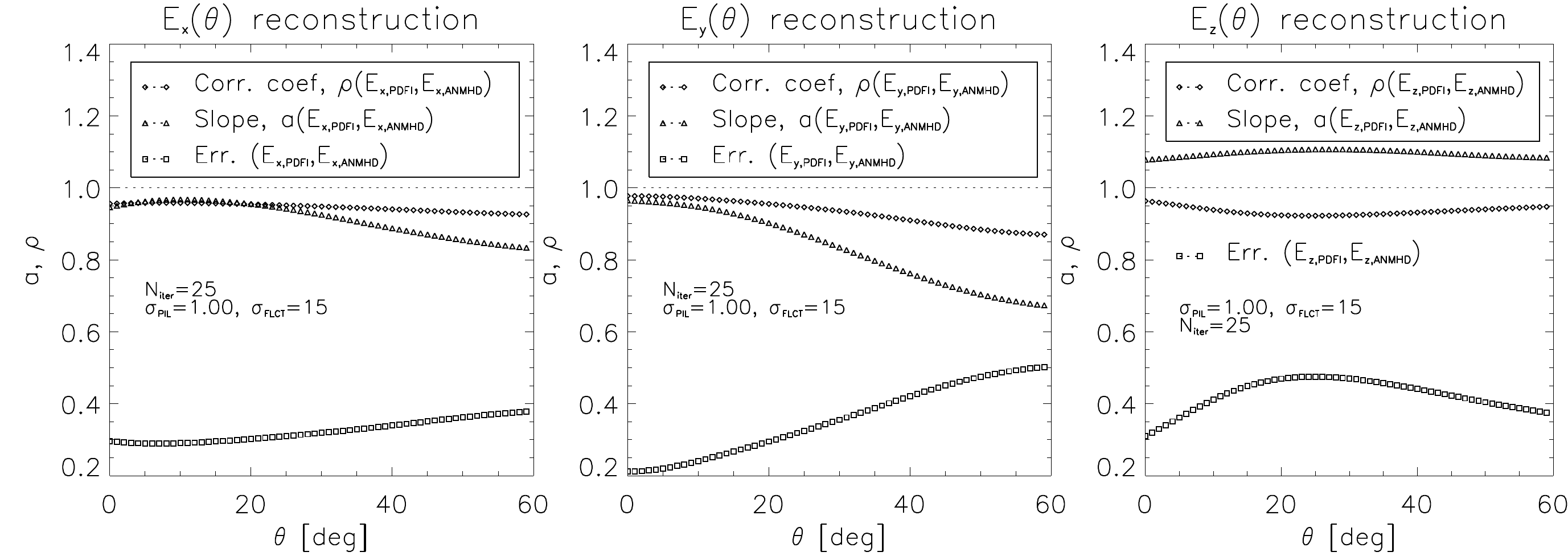}} 
  \caption{Validation of PDFI electric fields at viewing angles in the range of $\theta=[0^{\circ},60^{\circ}]$: quality of $E_x$ ({\it left}), $E_y$ ({\it middle}) and $E_z$ ({\it right}) reconstructions.}
  \label{fig_eang}
\end{figure*}

  \begin{figure*}[htb]
  \centering 
  \resizebox{1.0\hsize}{!}{\includegraphics[angle=0]{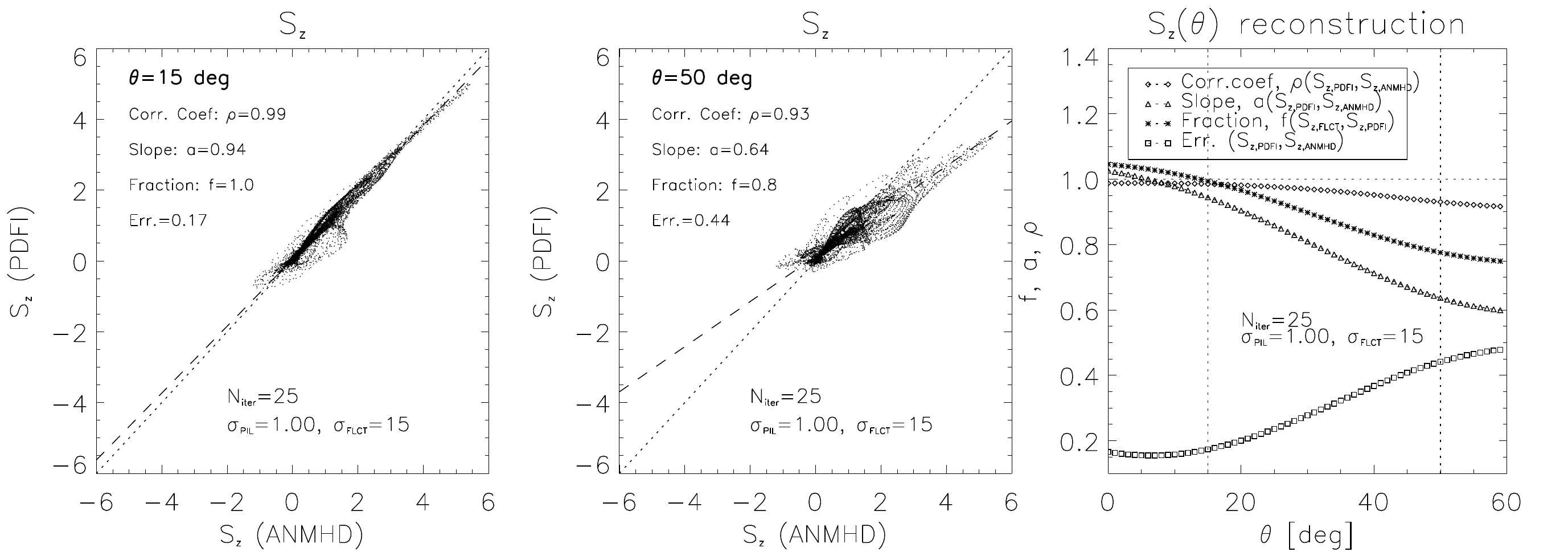}} 
  \caption{Validation of PDFI Poynting fluxes at viewing angles in the range of $\theta=[0^{\circ},60^{\circ}]$: {\it Left and Middle:} Pixel-to-pixel scatter plots comparing actual and PDFI Poynting fluxes at viewing angles $\theta=15^{\circ}$ ({\it left}) and $\theta=50^{\circ}$ ({\it right}) . {\it Right}: Quality of $S_z$ reconstructions in the range of $\theta=[0^{\circ},60^{\circ}]$ .}
  \label{fig_szang}
\end{figure*}

To test performance of PDFI method at non-normal viewing angles $\theta$, we calculated  the electric field (Figure~\ref{fig_eang}) and Poynting fluxes (Figure~\ref{fig_szang}) at values of  $\theta$ ranging from $0$ to $60$ degrees.  

Figure~\ref{fig_eang} shows that  at $\theta=0^\circ$ the error between reconstructed  and the actual \texttt{ANMHD} electric fields is the smallest, and as the angle increases, the quality of the reconstruction decreases.  For $\theta<30^{\circ},$ for all {\bf E}-components, PDFI correctly identifies the slope between reconstructed  and the actual \texttt{ANMHD} electric fields, within a $10\%$ difference.  At the largest angle, $\theta=60^\circ$, the slope is $a=0.83$ for $E_x$ and $a=0.67$ for $E_y$, and the error in these variables reaches up to $50\%$. In contrast to the horizontal electric field, the vertical component $E_z$ is relatively insensitive to the viewing angle: the slope $a$ varies within $[1.08,1.10]$. The latter is not surprising, since the Doppler contribution, which has most of the angular dependence, primarily constrains the horizontal field.  For all ${\bf E}$-components the correlation coefficient at $\theta=[0,60]^\circ$ is quite high, $\rho>0.9$, implying that the estimates of the slope shown on the plot adequately describe the quality of the reconstruction.  

Figure~\ref{fig_szang} describes the quality of the vertical Poynting flux reconstruction at different viewing angles.  Two left panels show a point-to-point comparison between the \texttt{ANMHD} and PDFI Poynting fluxes at $\theta=15^{\circ}$ (left panel) and $\theta=50^{\circ}$ (middle panel). At $\theta=15^{\circ}$  the agreement between the \texttt{ANMHD} and PDFI $S_z$ is very good: the slope $a=0.94, \rho=0.99, f=1, Err.=0.17$. Increasing the angle, at $\theta=50^{\circ}$ the scatter increases and the slope decreases to $a=0.64$, the error $Err.=0.44$ and PDFI method recovers $80\%$ of the total actual flux. The right panel summarizes the quality of the Poynting flux reconstruction for viewing angles within $[0,60^\circ]$-range.  At close-to-normal angles, $\theta<20^{\circ}$, PDFI recovers more than $95\%$ of the total energy flux and the error between reconstructed and the actual Poynting fluxes is less than 15\% ($Err.=0.15$). At larger angles, $\theta<60^\circ$, $75\%$ of the total flux is recovered and the error $Err.<0.5$. For the helicity flux rate, at $\theta<60^\circ$, more than $90\%$ of the total flux is recovered correctly and the error $Err.=0.3$.

\subsection{Comparison Of Electric Field Reconstruction Techniques}\label{anmhd_nind}

\begin{figure*}[htb]
  \centering 
  \resizebox{1.0\hsize}{!}{\includegraphics[angle=0]{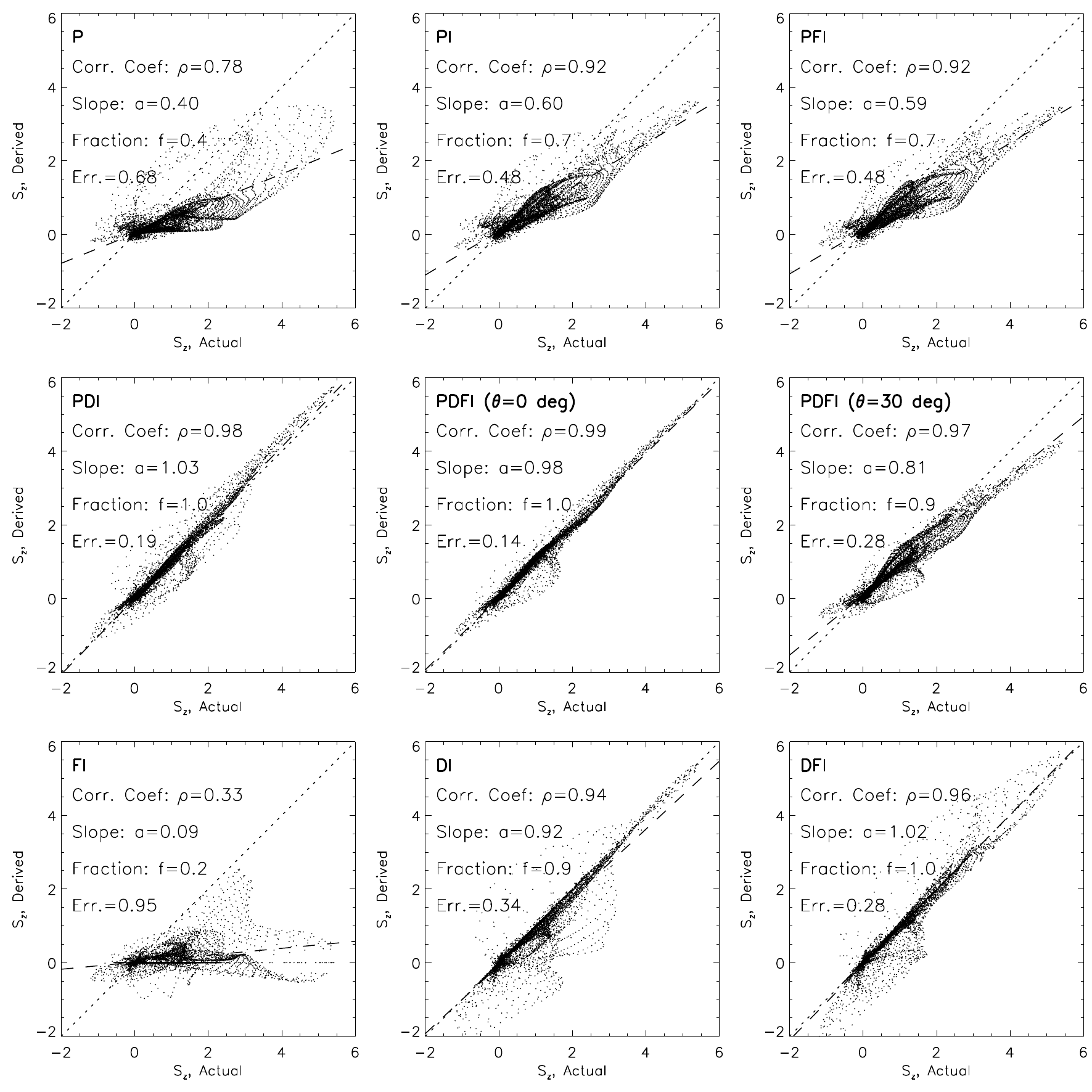}} 
  \caption{Comparison of the derived and the actual \texttt{ANMHD} vertical Poynting fluxes for different electric field inversion methods. Each method's results are shown in a separate panel with method's name indicated in the upper left corner and described in \S~\ref{esummary}. }
  \label{fig_szcmp}
\end{figure*}

\begin{figure*}[htb]
  \centering 
  \resizebox{1.0\hsize}{!}{\includegraphics[angle=0]{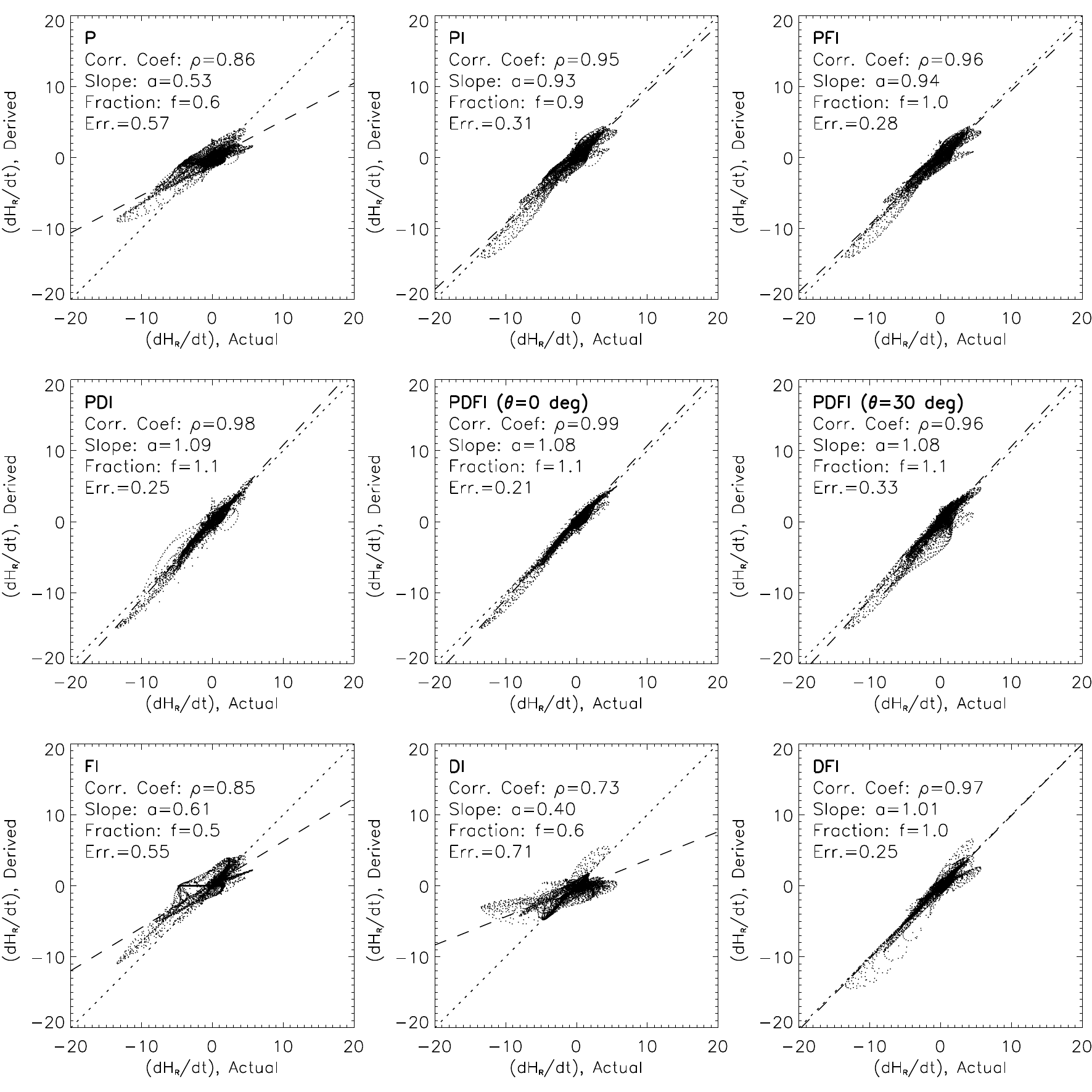}} 
  \caption{Comparison of the derived and the actual \texttt{ANMHD} helicity flux rates for different electric field inversion methods. Each method's results are shown in a separate panel with method's name indicated in the upper left corner and described in \S~\ref{esummary}. }
  \label{fig_helcmp}
\end{figure*}

\begin{figure*}[htb]
  \centering 
  \resizebox{1.0\hsize}{!}{\includegraphics[angle=0]{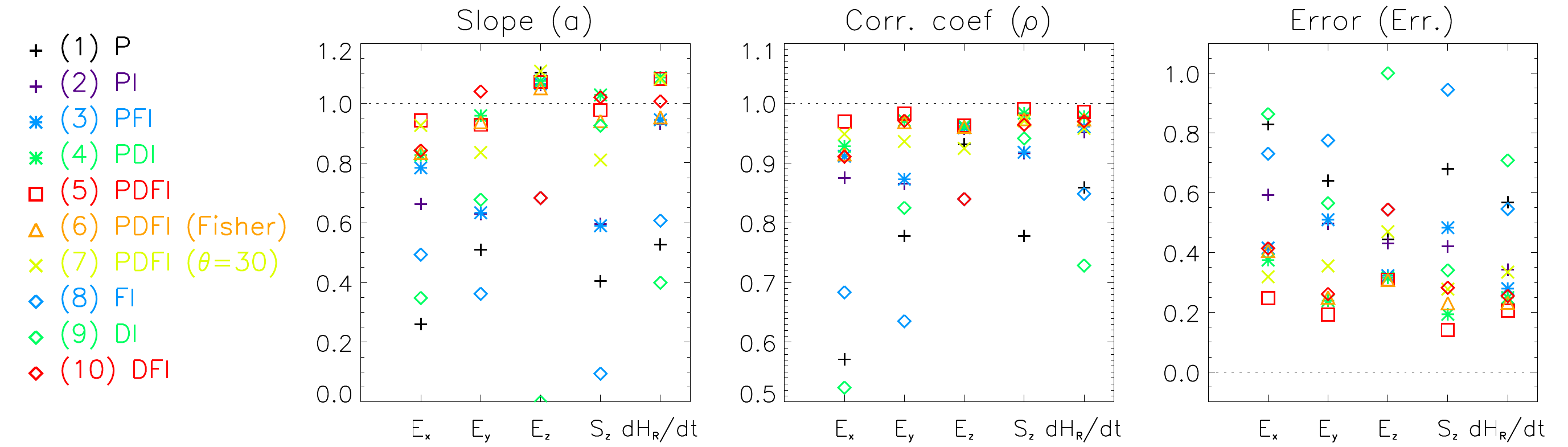}} 
  \caption{
  Comparison of $[E_x, E_y, E_z, S_z, \frac{dH_R}{dt}]$, derived using different electric field inversion techniques (1-10), with the actual \texttt{ANMHD} quantities, where $[E_x, E_y, E_z]$ are the three components of the electric field, $S_z$ is the vertical Poynting flux, $\frac{dH_R}{dt}$ is helicity flux rate. The methods we use for calculating electric fields (1-10) are in detail described in \S~\ref{esummary}.  {\it Left} panel shows slopes $a$ of the scatter-plots of reconstructed versus the actual \texttt{ANMHD} data.
 {\it Middle} panel shows correlation coefficients between the reconstructed and the actual data. {\it Right} panel shows errors $Err.=\sigma(u-u')/\sigma_{u}$, where $u$ and $u'$ are actual and derived values of one of the five analyzed quantities. Horizontal dotted lines correspond to the ideal reconstruction. The plotted values are given in Table~\ref{t1}. For description of $[a,\rho,Err.]$ see \S~\ref{anmhd_intro}. }
  \label{fig_cmp}
\end{figure*}

In this section we analyze roles that the inductive and non-inductive components play in reconstructed vertical Poynting flux (Figure~\ref{fig_szcmp}), electric field and helicity fluxes (Figures~\ref{fig_helcmp} and \ref{fig_cmp}, Table~\ref{t1}).
 
Figure~\ref{fig_szcmp} shows scatter plots comparing the actual vertical Poynting flux with the the Poynting fluxes derived with different reconstruction methods: (1) P, (2) PI, (3) PFI, (4) PDI, (5) PDFI, (6) PDFI at non-normal angle $\theta=30^{\circ}$, (7) FI, (8) DI, (9) DFI. The nomenclature that we use here is described in \S~\ref{esummary} and Table~\ref{table:esummary}. Using just the inductive part of the electric field (P), we reconstruct only $40\%$ of the total Poynting flux and the scatter from the linear dependence is large ($Err.=0.68$). Adding the ideal MHD assumption (PI) adds $30\%$ more Poynting flux, leading to much less scatter (larger $\rho$) and a smaller error ($Err.=0.48$). Inclusion of the non-ideal contribution due to horizontal plasma velocities, inferred from the \texttt{FLCT} (PFI), does not improve the solution. However, when we include the Doppler non-inductive component alone (PDI), we reconstruct $100\%$ of the flux, i.e. the role of the Doppler contribution is much higher than that from the \texttt{FLCT} horizontal velocities. Finally, when we add both the \texttt{FLCT} and the Doppler contributions (PDFI), the final Poynting flux is the closest to the actual $S_z$, with a slightly smaller error ($Err.=0.14$) than in the PDI case ($Err.=0.19$). On a separate note, if we use a non-PTD ideal inversion technique instead of the PTD,  ${\bf E}=-{\bf V} \times \bvec$, $V_z=0$  (FI), we reconstruct only $20\%$ of the flux, i.e. the agreement between the reconstruction and the actual $S_z$ is poor ($Err.=0.95$). If we know the vertical Doppler velocity (DI), then $90\%$ of the actual flux is reconstructed and with a good agreement ($\rho=0.94, a=0.92, f=0.9, Err.=0.34$). With both vertical and horizontal velocities (DFI), we extract $100\%$ of the total Poynting flux, the slope $a=1.02$ and the error $Err.=0.28$. How is this different from $S_z$ from the PDFI? The PDFI electric field that includes both inductive and non-inductive contributions yields a factor of two smaller error in $S_z$ ($Err.=0.14$) than the DFI field, and shows less scatter, especially in the regions of strong Poynting fluxes, thus it better represents the vertical Poynting flux. For completeness, in Figure~\ref{fig_helcmp} we also compare helicity flux rates, $\frac{dH_R}{dt}$, from different reconstruction methods with the actual helicity flux rate.  We remark that the crucial role the Doppler signal plays in reconstructing the Poynting and helicity fluxes in this case might be due to the process being modeled in the \texttt{ANMHD} simulation: an emerging magnetic bipole.  It is possible that Doppler inputs to electric field estimates are less important when flux is not actively emerging.

In Figure~\ref{fig_cmp} and Table~\ref{t1} we summarize the quality of different electric field inversion techniques in the same way we did for $S_z$ and $\frac{dH_R}{dt}$ for all the variables, $[E_x, E_y, E_z, S_z, \frac{dH_R}{dt}]$, where $[E_x, E_y, E_z]$ are the three components of the electric field, $S_z$ is the vertical Poynting flux,  $ \frac{dH_R}{dt}$ is the helicity flux rate (see Eq.~(\ref{eq_hele}-\ref{eq_helv})).  Besides the methods shown previously (see Figure~\ref{fig_szcmp}), we also add the method number 6, ``PDFI (Fisher)'', that compares variables reconstructed in \cite{Fisher2012} with the actual \texttt{ANMHD} values.

The P-solution yields the worst reconstructions -- the slope between the actual and derived $E_x$ and $E_y$ is $0.26$ and $0.51$ respectively and the correlation coefficient in both cases is less than $0.8$. The quantity $E_z$, however, is reconstructed very well -- the slope is $1.10$ and the correlation coefficient is $0.93$. Adding the other ingredients only slightly improves the slope and the correlation coefficient of $E_z$. The Poynting and helicity fluxes, $S_z$  and $\frac{dH_R}{dt}$ respectively, are reconstructed poorly: the slopes are $0.4$ and $0.53$ respectively.

Application of the ideal MHD constraint to the inductive solution (PI, case 2) significantly improves reconstruction of $E_x$ and $E_y$: the slope for $E_x$ increases from $0.26$ to $0.66$, and from $0.51$ to $0.63$ for $E_y$; this results in much better quality of the Poynting and helicity fluxes reconstructions: the slope for the Poynting flux changes from $0.4$ to $0.6$ from P- to PI-solution and for helicity flux - from $0.53$ to $0.93$; the error decreases by more than $20\%$.  

Addition of the non-inductive contribution from horizontal velocities (PFI, case 3), slightly improves the slope and correlation coefficient of $E_x$: the slope changes from $0.66$ to $0.78$ and the correlation coefficients - from $0.88$ to $0.91$.  The error decreases as well.

Inclusion of the non-inductive contribution from the vertical Doppler velocity, (PDI, case 4), provides much better improvement than the \texttt{FLCT} contribution (PFI) \citep{Fisher2012}. The slopes and correlation coefficients for both $E_x$ and $E_y$ increase significantly from $0.66$ and $0.63$ for the PI solution to $0.82$ and $0.96$, respectively. This results in much better reconstruction of $S_z$: the slope for $S_z$ changes from $0.6$ to $1.03$. The errors, especially for $E_y$, descrease, changing from $0.51$ to $0.24$. 

Finally, in the PDFI case (case 5) we add both non-inductive-, \texttt{FLCT} and Doppler  contributions. We find that the PDFI method yields the best agreement with the \texttt{ANMHD} variables. The slopes for $E_x$ and $E_y$ are $0.94$ and $0.93$ with the correlation coefficients of $0.97$ and $0.98$ respectively. The Poynting and helicity fluxes have slopes of $0.98$ and $1.08$ with correlation coefficients of $0.99$ and $0.99$. The error varies from $0.14$ (for $S_z$) to $0.31$ (for $E_z$).  To summarize, in terms of the fractions, PDFI method predicts roughly $100\%$ of the Poynting flux and $110\%$ of the helicity flux rate. It yields slopes and correlation coefficients that are similar to \cite{Fisher2012} (PDFI (Fisher), see case 6), but has smaller errors (compare $0.14$ vs. $0.23$ for $S_z$).

At a non-normal viewing angle of $\theta=30^{\circ}$ (case 7),  the agreement gets slightly worse, but still the slopes are close to one and the error, for example, in the Poynting flux is $28\%$. We reconstruct $90\%$ of the total Poynting flux and $110\%$ of the total helicity flux. Applying the same technique (\S~\ref{nonzero}) at normal viewing angle we find the same results as the ones derived using the normal viewing angle technique (\S~\ref{DEFzero}). 

We also calculate ${\bf E}, S_z$ and $dH_R/dt$ using the ideal non-PTD ${\bf E}=-{\bf V} \times \bvec$ formalism: the FI, DI and DFI (cases 8, 9 and 10). If only horizontal components of the velocity field are used and $V_z$ is set to zero, (FI), then we get the worst reconstruction -  the slopes for $E_x$ and $E_y$ are less than $0.5$.  The slope for the Poynting flux is $a=0.09$ and the fraction $f=0.5$. This is much worse than the PFI solution: $a=0.95$ and $f=1.0$. The difference between PFI- and FI-solutions might seem surprising at first, since both PFI and FI use the same information on the input. However while PFI solves the induction equation, the FI method does not. When we take the vertical velocity into account, (DI), the quantity $E_z$ is unknown, the quality of $E_x$ and $E_y$ improves slightly, raising from slopes of $a=0.49$ and $a=0.36$ to $a=0.35$ and $a=0.68$ respectively. As a result, the reconstruction of $S_z$ improves from $a=0.09$ to $a=0.92$ and the fraction is $f=0.9$. This is slightly worse than the  $S_z$ from PDI: $a=1.03$ and $f=1.0$. For the non-inductive field estimates, we get the best agreement for $S_z$, when both horizontal and Doppler velocities are taken into account (DFI, case 10): $a=1.02, f=1.0$ \citep[See also][]{Ravindra2008}. Comparing the DFI with the PDFI reconstruction, we find that  DFI  is able to capture $E_y$, Poynting and helicity fluxes, although with larger errors than the PDFI, while $E_x$ and $E_z$ yield a poor  reconstruction. In contrast, PDFI restores all three components of the electric field correctly, as well as Poynting and helicity fluxes, and with smaller errors. The reconstructions that include PTD (PI, PFI, PDI) also do a much better job than non-PTD solutions (FI, DI), when only one contribution to the velocity field (Doppler or FLCT) is available.

We separately compare actual and calculated helicity flux rates and estimate the quality of the helicity flux rate reconstruction when, instead of the electric field (PI, PDI, PFI), only velocity estimates are used (FI or DI,  see Eq.~(\ref{eq_helv})). When only the horizontal velocity field has been estimated (PFI or FI), then PFI yields much better reconstruction of helicity flux than FI: $a=0.95, f=1.0,Err.=0.28$ (PFI) vs. $a=0.61, f=0.5,Err.=0.55$ (FI). When only the Doppler velocity field is known (PDI or DI), then PDI also does a much better job than DI: $a=1.08, f=1.1, Err.=0.25$ (PDI) vs. $a=0.4,f=0.6, Err.=0.71$ (DI). Finally, if both horizontal and Doppler fields are known (PDFI or DFI), then both PDFI and DFI  get similar values of helicity fluxes. However, the PDFI solution gets slightly better correlation coefficients and smaller errors and hence less scatter.

Figure~\ref{fig_cmp} summarizes performance of all the methods in three panels (slope, correlation coefficient and error). For the  PDFI method, shown with red squares,  we get the most accurate electric fields, Poynting and helicity fluxes.  The PDFI method outperforms ideal non-PTD methods (FI, DI, DFI).   

 \renewcommand{\arraystretch}{1.6}
\renewcommand{\arraystretch}{1.6}
\begin{table*} [h] \caption{Comparison of $[E_x, E_y, E_z, S_z, \frac{dH_R}{dt}]$, derived using different electric field inversion techniques (1-10), with the actual \texttt{ANMHD} quantities. Each entry in the table has a form of   $\left[\hspace{2.5pt}^{\rho}\hspace{1pt}a\,_{Err.}\right]$, where slope $a$, correlation coefficient  $\rho$ and $Err.$ are described in \S~\ref{anmhd_intro}.  The ideal reconstruction satisfies $\left[\hspace{2.5pt}^{\rho}\hspace{1pt}a\,_{Err.}\right]=(^{1.00} 1.00 _{0.00})$. For a plot, see Figure~\ref{fig_cmp}.} 
\small \begin{center} \begin{tabular}{l|@{}c@{}@{}@{}c@{}@{}@{}c@{}@{}@{}c@{}@{}@{}c@{}@{}}
All            &$E_x$           &$E_y$           &$E_z$           &$S_z$           &$\left(\frac{dH_{R}}{dt} \right)$\\
\hline
(1) P  &\hspace{2.5pt}$^{0.57}$\hspace{1pt}0.26\,$_{0.83}$\hspace{3pt}
&\hspace{2.5pt}$^{0.78}$\hspace{1pt}0.51\,$_{0.64}$\hspace{3pt}
&\hspace{2.5pt}$^{0.93}$\hspace{1pt}1.10\,$_{0.44}$\hspace{3pt}
&\hspace{2.5pt}$^{0.78}$\hspace{1pt}0.40\,$_{0.68}$\hspace{3pt}
&\hspace{2.5pt}$^{0.86}$\hspace{1pt}0.53\,$_{0.57}$\hspace{3pt}
\\
(2) PI   &\hspace{2.5pt}$^{0.88}$\hspace{1pt}0.66\,$_{0.59}$\hspace{3pt}
&\hspace{2.5pt}$^{0.87}$\hspace{1pt}0.63\,$_{0.50}$\hspace{3pt}
&\hspace{2.5pt}$^{0.96}$\hspace{1pt}1.06\,$_{0.43}$\hspace{3pt}
&\hspace{2.5pt}$^{0.92}$\hspace{1pt}0.60\,$_{0.42}$\hspace{3pt}
&\hspace{2.5pt}$^{0.95}$\hspace{1pt}0.93\,$_{0.34}$\hspace{3pt}
\\
(3) PFI &\hspace{2.5pt}$^{0.91}$\hspace{1pt}0.78\,$_{0.42}$\hspace{3pt}
&\hspace{2.5pt}$^{0.87}$\hspace{1pt}0.63\,$_{0.51}$\hspace{3pt}
&\hspace{2.5pt}$^{0.96}$\hspace{1pt}1.06\,$_{0.32}$\hspace{3pt}
&\hspace{2.5pt}$^{0.92}$\hspace{1pt}0.59\,$_{0.48}$\hspace{3pt}
&\hspace{2.5pt}$^{0.96}$\hspace{1pt}0.94\,$_{0.28}$\hspace{3pt}
\\
(4) PDI &\hspace{2.5pt}$^{0.93}$\hspace{1pt}0.83\,$_{0.37}$\hspace{3pt}
&\hspace{2.5pt}$^{0.97}$\hspace{1pt}0.96\,$_{0.24}$\hspace{3pt}
&\hspace{2.5pt}$^{0.96}$\hspace{1pt}1.07\,$_{0.32}$\hspace{3pt}
&\hspace{2.5pt}$^{0.98}$\hspace{1pt}1.03\,$_{0.19}$\hspace{3pt}
&\hspace{2.5pt}$^{0.98}$\hspace{1pt}1.09\,$_{0.25}$\hspace{3pt}
\\
(5) PDFI &\hspace{2.5pt}$^{0.97}$\hspace{1pt}0.94\,$_{0.25}$\hspace{3pt}
&\hspace{2.5pt}$^{0.98}$\hspace{1pt}0.93\,$_{0.19}$\hspace{3pt}
&\hspace{2.5pt}$^{0.96}$\hspace{1pt}1.07\,$_{0.31}$\hspace{3pt}
&\hspace{2.5pt}$^{0.99}$\hspace{1pt}0.98\,$_{0.14}$\hspace{3pt}
&\hspace{2.5pt}$^{0.99}$\hspace{1pt}1.08\,$_{0.21}$\hspace{3pt}
\\
(6) PDFI (Fisher) &\hspace{2.5pt}$^{0.91}$\hspace{1pt}0.83\,$_{0.41}$\hspace{3pt}
&\hspace{2.5pt}$^{0.97}$\hspace{1pt}0.94\,$_{0.25}$\hspace{3pt}
&\hspace{2.5pt}$^{0.96}$\hspace{1pt}1.05\,$_{0.31}$\hspace{3pt}
&\hspace{2.5pt}$^{0.97}$\hspace{1pt}0.94\,$_{0.23}$\hspace{3pt}
&\hspace{2.5pt}$^{0.97}$\hspace{1pt}0.95\,$_{0.23}$\hspace{3pt}
\\
(7) PDFI ($\theta=30^\circ$) &\hspace{2.5pt}$^{0.95}$\hspace{1pt}0.93\,$_{0.32}$\hspace{3pt}
&\hspace{2.5pt}$^{0.94}$\hspace{1pt}0.83\,$_{0.36}$\hspace{3pt}
&\hspace{2.5pt}$^{0.92}$\hspace{1pt}1.11\,$_{0.47}$\hspace{3pt}
&\hspace{2.5pt}$^{0.97}$\hspace{1pt}0.81\,$_{0.28}$\hspace{3pt}
&\hspace{2.5pt}$^{0.96}$\hspace{1pt}1.08\,$_{0.33}$\hspace{3pt}
\\ \hline
(8) FI &\hspace{2.5pt}$^{0.68}$\hspace{1pt}0.49\,$_{0.73}$\hspace{3pt}
&\hspace{2.5pt}$^{0.64}$\hspace{1pt}0.36\,$_{0.78}$\hspace{3pt}
&\hspace{2.5pt}$^{0.84}$\hspace{1pt}0.68\,$_{0.54}$\hspace{3pt}
&\hspace{2.5pt}$^{0.33}$\hspace{1pt}0.09\,$_{0.95}$\hspace{3pt}
&\hspace{2.5pt}$^{0.85}$\hspace{1pt}0.61\,$_{0.55}$\hspace{3pt}
\\
(9) DI ($\theta=0^\circ$) &\hspace{2.5pt}$^{0.52}$\hspace{1pt}0.35\,$_{0.86}$\hspace{3pt}
&\hspace{2.5pt}$^{0.82}$\hspace{1pt}0.68\,$_{0.57}$\hspace{3pt}
&\hspace{2.5pt}$^{--}$\hspace{1pt}--\,$_{--}$\hspace{3pt}
&\hspace{2.5pt}$^{0.94}$\hspace{1pt}0.92\,$_{0.34}$\hspace{3pt}
&\hspace{2.5pt}$^{0.73}$\hspace{1pt}0.40\,$_{0.71}$\hspace{3pt}
\\
(10) DFI ($\theta=0^\circ$) &\hspace{2.5pt}$^{0.91}$\hspace{1pt}0.84\,$_{0.41}$\hspace{3pt}
&\hspace{2.5pt}$^{0.97}$\hspace{1pt}1.04\,$_{0.26}$\hspace{3pt}
&\hspace{2.5pt}$^{0.84}$\hspace{1pt}0.68\,$_{0.54}$\hspace{3pt}
&\hspace{2.5pt}$^{0.96}$\hspace{1pt}1.02\,$_{0.28}$\hspace{3pt}
&\hspace{2.5pt}$^{0.97}$\hspace{1pt}1.01\,$_{0.25}$\hspace{3pt}
\\
\end{tabular}\normalsize \label{t1} \end{center} \end{table*}







\section{DISCUSSION AND CONCLUSIONS}\label{disc}

We have modified the methods described in \cite{Fisher2012} for estimating electric fields from vector magnetic fields and Doppler velocities to incorporate non-normal viewing angles and the faster, more robust \texttt{FISHPACK} solver to produce a {\bf P}TD-{\bf D}oppler-{\bf F}LCT-{\bf I}deal-Ohm's solver (PDFI) that could be easily applied to observed data. We then used a pair of synthetic magnetograms extracted from MHD simulations, in which magnetic and Doppler velocity fields are known, to estimate the electric fields. Finally, we characterized the accuracy of the derived electric fields, as well as of the fluxes of magnetic energy and helicities.  In this section we summarize the strengths and weaknesses of the PDFI technique, compare its results with other techniques and describe how it will be used to analyze HMI vector magnetograms and Doppler data.

At zero viewing angle, the accuracy of the PDFI method is excellent. Using the PDFI method, the total Poynting flux, $S_z$, is estimated with an error of less than $1\%$ and the total helicity flux rate, $\frac{dH_R}{dt}$, with an error of $10\%$. The RMS of the integrands are $12\%$ and $26\%$ respectively.  These estimates are more accurate, and much faster to derive than those from \cite{Fisher2012} (see Figures~\ref{fig_sz0} and \ref{fig_hel0} and Table~\ref{t1}).\footnote{For the Poynting flux the slope between the actual and the PDFI $S_z$ is $a=0.98$ in this paper versus $a=0.94$ in \cite{Fisher2012} and  for the helicity flux rate, the slope $a=1.08$ in this paper versus $a=0.95$ in \cite{Fisher2012}; in both cases the scatter in the PDFIs is smaller than in \cite{Fisher2012}.} 

With increasing viewing angle $\theta$, the accuracy of the PDFI method slowly decreases. At  $\theta<20^{\circ}$ PDFI recovers more than $95\%$ of the total Poynting flux with a slope of $a=0.98$. At $\theta<60^\circ$ PDFI recovers more than $75\%$ of $S_z$ with the slope of $a=0.6$  (see Figure~\ref{fig_szang}). As for the helicity flux rate, at  $\theta<60^\circ$ more than $90\%$ of the total $\frac{dH_R}{dt}$ is recovered with the slope of $a=1.0$.

In this paper we compare the quality of reconstructions of electric fields, Poynting and helicity fluxes from the PTD-based methods (PFI, PDI, PDFI), that explicitly enforce consistency of electric fields with evolution of $\vecB$, with non-PTD methods (FI, DI, DFI), like e.g. FLCT, that derive horizontal velocities by tracking the vertical magnetic field. We show that when both horizontal and Doppler velocity fields are known, the two approaches, PDFI and DFI, yield similar results for helicity and Poynting fluxes, with PTD-based PDFI method having slightly smaller errors. However, when either the Doppler or the FLCT velocity field is unknown, the PTD methods  are much better in reconstructing both helicity and Poynting fluxes  than non-PTD methods.  For example, when only the horizontal velocity is known  PFI reconstructs $70\%$ of total $S_z$ and $100\%$ of total $\frac{dH_R}{dt}$ while FI reconstruct only $20\%$ and $50\%$ respectively.  Similarly, when only the Doppler velocity is known, then PDI reconstructs $100\%$ of total $S_z$ and $110\%$ of total $\frac{dH_R}{dt}$, while DI reconstructs  $90\%$ and $60\%$ respectively. 

For the helicity flux rate, our results imply, that in order to correctly capture the helicity flux rate, one cannot only use the velocity field determined from the tracking of the vertical magnetic field (FI), but has to separately include the emergence term due to vertical velocity:  the \cite{Demoulin2003} conjecture does not apply here, consistent with conclusions of \cite{Schuck2008,Liu2012,Ravindra2008}.
We note, again, that the \texttt{ANMHD} data we analyze were drawn from a simulation of an emerging magnetic bipole, a configuration in which vertical flows (used as our Doppler velocity input) play particularly strong roles in the fluxes of magnetic energy and helicity.  This might not be true in active regions when substantial amounts of new flux are not emerging.

To put our results into context, in Table~\ref{table:comp_sz} we compare PDFI Poynting and helicity fluxes that we derive in this paper to those calculated with \texttt{DAVE4VM} and \texttt{DAVE+ANMHD} \citep{Schuck2008}. While \texttt{DAVE4VM}  predicts roughly $75\%$ of the total Poynting flux and $95\%$ of the helicity flux rate, the PDFI method has a better performance, with less than  $1\%$ error in the total Poynting flux and a $10\%$ error in the helicity flux rate. One should keep in mind, however, that \texttt{DAVE4VM}, unlike PDFI, does not take the Doppler velocity into account, hence it is fair to compare not the PDFI's, but the PFI's and \texttt{DAVE4VM}'s results.  When we do that we find that the PFI performs similarly to \texttt{DAVE4VM}, and that both miss more than $25\%$ of the total Poynting flux.  In contrast, PDFI, which includes both horizontal and Doppler velocities, yields an excellent agreement  with the actual \texttt{ANMHD} quantities (see Table~\ref{table:comp_sz}). Note that the PDFI estimate for helicity flux is  better than that of the \texttt{DAVE+ANMHD}, that takes the Doppler signal into account (see Fig. 14 in \cite{Schuck2008}).  In addition, we remark that while the MEF method \citep{Longcope2004c} performed well in the tests by \cite{Welsch2007}, here we find the PDFI method to be superior, by several statistical measures, for the same test data (compare Figure 14 in \cite{Welsch2007} with Figure~\ref{fig_szcmp}).

\begin{table*}[h] \caption{Comparison of accuracy of Poynting and rate of relative helicity fluxes estimates between the PDFI, PFI, \texttt{DAVE+ANMHD} and \texttt{DAVE4VM} \citep{Schuck2008} over $|B_z|>370G$. An ideal reconstruction satisfies $a=1, \rho=1, f=1$.}
\small \begin{center} \begin{tabular}{p{2.4cm}|p{1.2cm}p{1.2cm}p{1.2cm}p{2.0cm}||p{1.2cm}p{1.2cm}p{1.2cm}p{2.0cm}}
 &PFI&\texttt{DAVE4VM} &PDFI & \small{\texttt{DAVE+ANMHD}} &PFI&\texttt{DAVE4VM} &PDFI & \texttt{DAVE+ANMHD} \\\hline
& \multicolumn{4}{ c|| }{Poynting flux, $S_z$} & \multicolumn{4}{ c }{Helicity flux rate, $\frac{dH_R}{dt}$}\\ 
Slope, $a$ &  $0.92$ & $0.71$ & $0.99$ & $0.99$ &  $0.96$ & $0.9$ & $0.99$ & $1.16$ \\ 
Corr. coef., $\rho$ &  $0.59$ & $0.83$ & $0.98$ & $0.96$  &  $0.95$ & $0.94$ & $1.08$ & $0.96$ \\ 
Fraction, $f$ &  $0.7$ & $0.76$ & $1.0$ & $0.99$ &  $1.0$ & $0.94$ & $1.1$ & $1.46$ \\ 
\end{tabular} \label{table:comp_sz} \end{center} \end{table*}

To estimate the speed of the PDFI method we did a series of inversion runs for the \texttt{ANMHD} dataset on a MacBook Pro laptop with 2GHz Intel Core i7 Processor and 8 GB 1333 MHz DDR3. For this test case, where $N_x=288$ and $N_y=288$ pixels,  it takes $0.24$ seconds to calculate the plain P electric field, $1.2$ seconds for the PDF electric field and $7.3$ seconds for the ideal PDFI electric field ($N_{iter}=25$, see Figure~\ref{fig_iter}).

One of the major weaknesses of the PDFI method, and also of any technique that uses the Doppler data, is a strong dependence on the Doppler bias velocity (see \S~\ref{anmhd_esh}). In this paper, to calculate PDFI electric fields, we assumed that we know the Doppler velocity field. In reality, however, the observed LOS velocity has a bias due to instrumental variations and the known correlation between the intensity and blue-shift, known as convective blueshift. Recently, \cite{Welsch2013} addressed this issue by presenting several methods to estimate the absolute calibration of LOS velocities in solar active regions near disk center that we apply to the HMI observations of the active region NOAA 11158 \citep{Kazachenko2014b}.

Another weakness of the PDFI method is the absence of clear understanding of the extent to which additional information from other sources (besides Doppler and horizontal velocities) is necessary to fully specify the non-inductive part of the electric field. In the past \cite{Fisher2012} showed that the Doppler signal near PILs and horizontal velocities away from the PIL contain important non-inductive information for the electric field that cannot be derived from the Faraday's law. In this paper, we tested this approach at non-zero viewing angles and found a good agreement for viewing angles less than $50^\circ$ for the \texttt{ANMHD} test case. There may be additional degrees of freedom for the non-inductive electric field which are not fully captured by the Doppler plus transverse magnetic field flux-emergence contribution described in \cite{Fisher2012}.  For example, rigid rotations of magnetic structures (e.g. sunspot rotation) that include a high degree of symmetry will have electric field components that are not fully captured by the PDFI formalism, because of the lack of change they produce in the vector magnetic field on the photosphere. Yet such motions transport significant energy and helicity into the corona \citep{Kazachenko2009}. For this reason, other tests of the emergence of a twisted flux tube from the interior into the solar atmosphere with different subsurface twisted flux tube configurations should be done. These tests will allow us formulate and then to incorporate any necessary additional electric fields corresponding to such horizontal vortical motions into the PDFI solution using observational information, such as observed sunspot penumbral motions as input. Furthermore, Doppler observations from a non-normal viewing angle can also be used to capture such horizontal motions, and help to recover the corresponding electric fields.

\acknowledgments
We thank the US taxpayers for providing the funding that made this research possible.
We thank Mark Cheung and the anonymous referee for their thoughtful input that have improved
the manuscript. We acknowledge funding from the Coronal Global Evolutionary Model (CGEM) award NSF AGS 1321474 (MDK, BTW, GHF),  NSF Award AGS-1048318 (GHF), NASA Award NNX13AK54G (MDK), NSF SHINE Postdoc Award 1027296 (MDK),  NSF's National Space Weather Program AGS-1024862 (BTW), the NASA Living-With-a-Star TR\&T Program NNX11AQ56G (MDK, BTW, GHF), and the NASA Heliophysics Theory Program NNX11AJ65G (GHF,  BTW).

\appendix
\section{Generalization Of PTD Formalism to Spherical Coordinates}\label{sphere}

We can extend the PTD formalism from Cartesian to spherical coordinates and show
that the form of the equations we need to solve to find the electric field vector is the same in both coordinate systems.

Similar to the Cartesian case (see Equations~(\ref{eq1}-\ref{eq4})), we decompose the magnetic field vector 
${\bvec}=(B_r,B_{\theta},B_{\phi})$  into two ``poloidal'' and  ``toroidal''  potentials, $\mathcal{B}$ and $\mathcal{J}$: $\bvec=\nabla \times \nabla \times {\mathcal{B}} 
{\bf \hat{r}} + \nabla \times {\mathcal{J}} {\bf \hat{r}}$ \citep{Moffatt1978}. We then take a partial time derivative of $\bvec$ and express the $(\nabla \times)$-operator  in spherical coordinates:
\begin{eqnarray*}
\dot{ {\bvec}}
&=&\nabla \times \nabla \times \dot{\mathcal{B}} {\bf \hat{r}} 
+ \nabla \times \dot{\mathcal{J}} {\bf \hat{r}} \\
&=&-\nabla^2 \dot{\mathcal{B}}{\bf \hat{r}} 
+ \nabla (\nabla \cdot \dot{\mathcal{B}} {\bf \hat{r}}) 
+ \nabla \times \dot{\mathcal{J}} {\bf \hat{r}} \\
&=&
\underbrace{\left[
\frac{-1}{r^2 \sin{\theta}} \frac{\partial}{\partial \theta }
\left(\sin\theta \frac{\partial \dot{\mathcal{B}}}{\partial \theta} \right )-
\frac{1}{r^2 \sin^2{\theta}}
\left(\frac{\partial^2 \dot{\mathcal{B}}}{\partial \phi^2}\right)
\right]}_{-\nabla^2_h \dot{\mathcal{B}}} \hat{r}+ \label{a_eq1} \\
&&+
\left[\frac{1}{r} 
\frac{\partial}{\partial \theta} 
\left(\frac{\partial \dot{\mathcal{B}}}{\partial r}\right)\right]
\hat{\theta}+
\left[\frac{1}{r \sin\theta} \frac{\partial}{\partial \phi} 
\left(\frac{\partial \dot{\mathcal{B}}}{\partial r}\right)\right]\hat{\phi}
+\nabla \times \dot{\mathcal{J}} \hat{r}\\
&=&-\nabla_h^2\dot{\mathcal{B}}\hat{r}+
\vec{\nabla}_h\left(\frac{\partial \dot{\mathcal{B}}}{\partial r}\right)
+
\nabla \times \dot{\mathcal{J}} \hat{r} \label{a_eq2}.
\end{eqnarray*}

By examining the $r$-component of the equation above, its horizontal divergence,
and the r-component of its curl,
we find three two-dimensional Poisson 
equations for the unknown functions $\dot{\mathcal{B}}$, 
$\frac{\partial \dot \mathcal{B}}{\partial z}$ and  $\dot{\mathcal{J}}$:
\begin{equation}
 -\dot{B_r}=\nabla^2 _h\dot{\mathcal{B}}, 
\label{a_eq3}
\end{equation}
\begin{equation}
\nabla_h \cdot \dot{\vecB}_h = \nabla^2_h 
\left(  \frac{\partial \dot \mathcal{B}}{\partial r} \right)
\label{a_eq41}
\end{equation}
\begin{equation}
- \hat{r} \cdot \left( \nabla \times \dot{\vecB}_h \right) =\nabla^2 _h\dot{\mathcal{J}},
\label{a_eq4}
\end{equation}
where the spherical form for the horizontal component of Laplace operator, 
$\nabla^2 _h$, is shown above and the horizontal divergence operator $(\nabla_h \cdot )$ has the standard expression for the divergence in spherical coordinates, without the $r$-derivative term.  Equations~(\ref{a_eq3}-\ref{a_eq4}) can be solved with the \texttt{HWSSSP} subroutine in \texttt{FISHPACK}  \citep{Schwarztrauber1974a}.

Uncurling Equation (\ref{a_eq2}) and comparing it with the Faraday's Law, we derive the electric field 
${\evec}$ in terms of calculated $\dot{\mathcal{B}}$ and  $\dot{\mathcal{J}}$ in spherical coordinates:
\begin{equation}
 c  {\evec} = c(E_r,E_{\phi},E_{\theta})=
- \nabla \times \dot{\mathcal{B}} {\bf \hat{r}} 
-\dot{\mathcal{J}} {\bf \hat{r}}-\nabla \psi 
\equiv  {c {\bf E^{P}}}-{\nabla \psi .}
\label{a_eq7}
\end{equation}

\clearpage
\bibliographystyle{kluwotitles}

\end{document}